%% file: compiler.tex
\documentclass[format=acmsmall, screen=true]{acmart}

\usepackage{booktabs} 

\usepackage{caption} 
\usepackage{subcaption} 

\usepackage{listings}  
\newfloat{listing}{thbp}{lst}
\floatname{listing}{Listing}
\usepackage{xspace}  

\usepackage[algo2e,ruled,longend,linesnumbered]{algorithm2e}
\SetAlFnt{\footnotesize}
\SetAlCapFnt{\footnotesize}
\SetAlCapNameFnt{\footnotesize}
\SetAlCapHSkip{0pt}
\IncMargin{-\parindent}
\SetKwInput{kwInput}{Input}
\SetKwInput{kwOutput}{Output}
\SetKwComment{Comment}{// }{}
\SetKw{Break}{break }{}
\SetKw{PyFor}{for}
\SetKwBlock{Try}{try}{}
\SetKwBlock{Except}{except}{}
\SetKw{Raise}{raise}
\SetKw{PyIf}{if}
\SetKw{Not}{not}
\SetKw{Or}{or}
\SetKw{And}{and}
\SetKw{In}{in}
\SetKw{Is}{is}
\SetKw{True}{true}
\SetKw{False}{false}
\SetKw{Skip}{skip}
\SetKwRepeat{Do}{do}{while}
\SetKwComment{Comment}{// }{}
\let\oldnl\nl
\newcommand{\nonl}{\renewcommand{\nl}{\let\nl\oldnl}}

\usepackage{tabulary} 

\usepackage{multicol}
\usepackage{multirow}
\usepackage{boldline}
\usepackage{amssymb}
\usepackage{enumitem}
\setlist[itemize]{leftmargin=5mm} 

\usepackage{array}
\newcolumntype{P}[1]{>{\centering\arraybackslash}p{#1}}
\newcolumntype{M}[1]{>{\centering\arraybackslash}m{#1}}
\newcolumntype{N}{@{}m{0pt}@{}}

\usepackage{mathtools}
\DeclarePairedDelimiter{\floor}{\lfloor}{\rfloor}

\newcommand{\Intel}{{Intel\textsuperscript{\tiny{\textregistered}}}\xspace}
\newcommand{\Xeon}{{Xeon\textsuperscript{\tiny{\textregistered}}}\xspace}
\newcommand{\Skylake}{{\Intel \xspace \Xeon Platinum 8180 (formerly code-named Skylake)}\xspace}
\newcommand{\XeonPhi}{{\Intel \xspace Xeon Phi\textsuperscript{\tiny{TM}}} \xspace}
\newcommand{\KNL}{{\XeonPhi 7250 (formerly code-named Knights Landing)}\xspace}
\newcommand{\vtune}{{\Intel \xspace VTune\textsuperscript{\tiny{\texttrademark}}}\xspace}
\newcommand{\Sympy}{{\em SymPy}\xspace}
\newcommand{\numpy}{{\em NumPy}\xspace}
\newcommand{\Python}{{\em Python}\xspace}

\newcommand{\CGen}{{\em CGen}\xspace}
\newcommand{\Grid}{{\tt Grid}\xspace}
\newcommand{\Function}{{\tt Function}\xspace}
\newcommand{\TimeFunction}{{\tt Time\-Function}\xspace}
\newcommand{\SparseFunction}{{\tt SparseFunction}\xspace}
\newcommand{\SparseTimeFunction}{{\tt Sparse\-Time\-Function}\xspace}
\newcommand{\PrecomputedSparseFunction}{{\tt Precomputed\-Sparse\-Function}\xspace}
\newcommand{\Dimension}{{\tt Dimension}\xspace}
\newcommand{\ConditionalDimension}{{\tt Conditional\-Dimen\-sion}\xspace}

\newcommand{\Operator}{{\tt Operator}\xspace}

\newcommand{\Space}{{\tt Space}\xspace}
\newcommand{\ISpace}{{\tt ISpace}\xspace}
\newcommand{\DSpace}{{\tt DSpace}\xspace}
\newcommand{\Cluster}{{\tt Cluster}\xspace}
\newcommand{\Iteration}{{\tt Itera\-tion}\xspace}
\newcommand{\Expression}{{\tt Expression}\xspace}

\newcommand{\skl}{{\tt skl8180}\xspace}
\newcommand{\knl}{{\tt knl7250}\xspace}
\newcommand{\isotropic}{{\tt isotro\-pic}\xspace}
\newcommand{\tti}{{\tt tti}\xspace}
\newcommand{\core}{{\tt core}\xspace}
\newcommand{\ops}{{\tt ops}\xspace}
\newcommand{\yask}{{\tt yask}\xspace}
\newcommand{\DevitoVer}{{\em Devito v3.1}\xspace}
\newcommand{\DevitoVerNext}{{\em Devito v3.2}\xspace}
\newcommand{\optbasic}{{\tt basic}\xspace}
\newcommand{\optadv}{{\tt advanced}\xspace}
\newcommand{\optagg}{{\tt aggressive}\xspace}
\newcommand{\shape}{{\tt shape}\xspace}
\newcommand{\extent}{{\tt extent}\xspace}
\newcommand{\origin}{{\tt origin}\xspace}
\newcommand{\name}{{\tt name}\xspace}
\newcommand{\data}{{\tt data}\xspace}
\newcommand{\coordinates}{{\tt coordinates}\xspace}

\newcommand{\so}{{\tt so}\xspace}
\newcommand{\FD}{FD\xspace}

\usepackage{fancybox}
\makeatletter
\newenvironment{CenteredBox}{%
\begin{Sbox}}{
\end{Sbox}\centerline{\parbox{\wd\@Sbox}{\TheSbox}}}
\makeatother

\begin{document}

\title[Devito, a system for automated stencil computation]{Architecture and performance of Devito, a system for automated stencil
computation}


\author{Fabio Luporini}
\affiliation{%
    \institution{Imperial College London}
}
\email{f.luporini12@imperial.ac.uk}

\author{Mathias Louboutin}
\affiliation{%
    \institution{Georgia Institute of Technology}
}
\email{mlouboutin3@gatech.edu}

\author{Michael Lange}
\affiliation{%
    \institution{European Centre for Medium-Range Weather Forecasts}
}
\email{michael.lange@ecmwf.int}

\author{Navjot Kukreja}
\affiliation{%
    \institution{Imperial College London}
}
\email{n.kukreja@imperial.ac.uk}

\author{Philipp Witte}
\affiliation{%
    \institution{Georgia Institute of Technology}
}
\email{pwitte3@gatech.edu}

\author{Jan H\"uckelheim}
\affiliation{%
    \institution{Imperial College London}
}
\email{j.hueckelheim@imperial.ac.uk}

\author{Charles Yount}
\affiliation{%
    \institution{Intel Corporation}
}
\email{chuck.yount@intel.com}

\author{Paul H. J. Kelly}
\affiliation{%
    \institution{Imperial College London}
}
\email{p.kelly@imperial.ac.uk}

\author{Felix J. Herrmann}
\affiliation{%
    \institution{Georgia Institute of Technology}
}
\email{felix.herrmann@gatech.edu}

\author{Gerard J. Gorman}
\affiliation{%
    \institution{Imperial College London}
}
\email{g.gorman@imperial.ac.uk}


\begin{abstract}
Stencil computations are a key part of many high-performance computing
applications, such as image processing, convolutional neural networks, and
finite-difference solvers for partial differential equations. Devito is a
framework capable of generating highly-optimized code given symbolic
equations expressed in \Python, specialized in, but not limited to, affine
(stencil) codes. The lowering process---from mathematical equations down
to C++ code---is performed by the Devito compiler through a series of
intermediate representations. Several performance optimizations are
introduced, including advanced common sub-expressions elimination, tiling
and parallelization. Some of these are obtained through well-established
stencil optimizers, integrated in the back-end of the Devito compiler. The
architecture of the Devito compiler, as well as the performance
optimizations that are applied when generating code, are presented. The
effectiveness of such performance optimizations is demonstrated using
operators drawn from seismic imaging applications.
\end{abstract}

%
%

%
%


\thanks{This work was supported by the Engineering and Physical Sciences
Research Council through grants EP/I00677X/1, EP/L000407/1, EP/I012036/1], by
the Imperial College London Department of Computing, by the Imperial College
London Intel Parallel Computing Centre (IPCC), and by the U.S. Department of
Energy, Office of Science, Office of Advanced Scientific Computing Research,
Applied Mathematics and Computer Science programs under contract number
DE-AC02-06CH11357.}

\maketitle

\renewcommand{\shortauthors}{F. Luporini et al.}

\input{body}

\bibliographystyle{ACM-Reference-Format}
\bibliography{bibliography}

\end{document}

%% file: body.tex
\section{Introduction}
Developing software for high-performance computing requires a considerable
interdisciplinary effort, as it often involves domain knowledge from numerous
fields such as physics, numerical analysis, software engineering, and low-level
performance optimization. The result is typically a monolithic application
where hardware-specific optimizations, numerical methods, and physical
approximations are interwoven and dispersed throughout a large number of loops,
functions, files, and modules. This frequently leads to slow innovation, high
maintenance costs, and code that is hard to debug and port onto new computer
architectures. A powerful approach to alleviate this problem is to introduce a
separation of concerns and to raise the level of abstraction by using
domain-specific languages (DSLs). DSLs can be used to express numerical methods
using a syntax that closely mirrors how they are expressed mathematically,
while a stack of compilers and libraries is responsible for automatically
creating the optimized low-level implementation in a general purpose
programming language such as C++.  While the focus of this paper is on
finite-difference (FD) based codes, the DSL approach has already had remarkable
success in other numerical methods such as the finite-element (FE) and
finite-volume (FV) method, as documented in Section~\ref{sec:related-work}.

This work describes the architecture of {\it Devito}, a system for automated
stencil computations from a high-level mathematical syntax. Devito was
developed with an emphasis on FD methods on structured grids. For this reason,
Devito's underlying DSL has many features to simplify the specification of FD
methods, as discussed in Section~\ref{sec:api}. The original motivation was to
solve large-scale partial differential equations (PDEs) in the context of
seismic inverse problems, where FD methods are commonly used for solving wave
equations as part of complex workflows (e.g., data inversion using
adjoint-state methods and backpropogation). Devito is equally useful as a
framework for other stencil computations in general; for example, computations
where all array indices are affine functions of loop variables. The Devito
compiler is also capable of generating arbitrarily nested, possibly irregular,
loops. This key feature is needed to support many complex algorithms that are
used in engineering and scientific practice, including applications from image
processing, cellular automata, and machine-learning.

One of the design goals of Devito was to enable high-productivity, so it is
fully written in \Python, with easy access to solvers, optimizers, input and
output, and the wide range of other libraries in the \Python ecosystem. At the
same time, Devito transforms high-level symbolic input into optimized C++ code,
resulting in a performance that is competitive with hand-optimized
implementations.  While the examples presented in this paper focus on using
Devito from a \Python application, exploiting the full potential of on-the-fly
code generation and just-in-time (JIT) compilation, a practical advantage of
generating C++ as an intermediate step is that it can be also used to generate
libraries for legacy software, thus enabling incremental code modernization.

Compared to other DSL frameworks that are used in practice, Devito uses
compiler technology, including several layers of intermediate representations,
to perform optimizations in multiple passes. This allows Devito to perform more
complex optimizations and to better optimize the code for individual target
platforms.  The fact that these optimizations are performed programmatically
facilitates performance portability across different computer
architectures~\cite{pennycook2016metric}. This is important, as industrial
codes are often used on a variety of platforms, including clusters with
multi-core CPUs, GPUs, and many-core chips spread across several compute nodes
as well as various cloud platforms. Devito also performs high-level
transformations for floating-point operation (FLOP) reduction based on symbolic
manipulation, as well as loop-level optimizations as implemented in Devito's
own optimizer, or using a third-party stencil compiler such as
YASK~\cite{yask-main}. The Devito compiler is presented in detail in
Sections~\ref{sec:compiler}, \ref{sec:optimizations}, and \ref{sec:sie}.

After the presentation of the Devito compiler, we show test cases in
Section~\ref{sec:performance} that are inspired by real-world seismic-imaging
problems.  The paper finishes with directions for future work and conclusions in
Sections~\ref{sec:further-work} and \ref{sec:conclusions}.


\section{Related work}
\label{sec:related-work}
The objective of maximizing productivity and performance through frameworks
based upon DSLs has long been pursued. In addition to well-known systems such
as Mathematica\textsuperscript{\tiny{\textregistered}} and
Matlab\textsuperscript{\tiny{\textregistered}}, which span broad mathematical
areas, there is a number of tools specialized in numerical methods for PDEs,
some dating back to the 1970s~\cite{PDEL,deqsol,alpal,ctadel}.

\subsection{DSL-based frameworks for partial differential equations}
One noteworthy contemporary framework centered on DSLs is FEniCS~\cite{Fenics},
which allows the specification of weak variational forms, via UFL~\cite{UFL},
and finite-element methods, through a high-level syntax.
Firedrake~\cite{Firedrake} implements the same languages as FEniCS, although it
differs from it in a number of features and architectural choices. Devito is
heavily influenced by these two successful projects, in particular by their
philosophy and design. Since solving a PDE is often a small step of a larger
workflow, the choice of \Python to implement this software provides access to
a wide ecosystem of scientific packages. Firedrake also follows the principle
of graceful degradation, by providing a very simple lower-level API to escape
the abstraction when non-standard calculations (i.e., unrelated to the
finite-element formulation) are required. Likewise, Devito allows injecting
arbitrary expressions into the finite-difference specification; this feature
has been used in real-life cases, for example for interpolation in seismic
imaging operators. On the other hand, a major difference is that Devito lacks a
formal specification language such us UFL in FEniCS/Firedrake. This is partly
because there is no systematic foundation underpinning FD, as opposed to FE
which relies upon the theory of Hilbert spaces~\cite{fem-scott}. Yet another
distinction is that, for performance reasons, Devito takes control of the
time-stepping loop. Other examples of embedded DSLs are provided by the
OpenFOAM project, with a language for FV~\cite{OpenFOAM}, and by PyFR, which
targets flux reconstruction methods~\cite{PyFR}.

\subsection{High-level approaches to finite differences}
Due to its simplicity, the FD method has been the subject of multiple research
projects, chiefly targeting the design of effective software abstraction and/or
the generation of high performance code
~\cite{STARGATES,simflowny,OpenSBLI,ExaStencils}. Devito distinguishes itself
from previous work in a number of ways, including support for the principle of
graceful degradation for when the DSL does not cover a feature required by an
application; incorporation of a symbolic mathematics engine; using actual
compiler technology rather than template-based code generation; and adoption of a
native \Python interface that naturally allows composition into complex
workflows such as optimization and machine-learning frameworks.

At a lower level of abstraction there are a number of tools targeting ``stencil''
computation (FD codes belong to this class), whose major objective is the
generation of efficient code. Some of them provide a
DSL~\cite{yask-main,OPS,Zhang-stencil,Halide}, whereas others are compilers or
user-driven code generation systems, often based upon a polyhedral model, such
as~\cite{pluto,loopy}. From the Devito standpoint, the aim is to harness these
tools---for example by integrating them---to maximize performance
portability. As a proof of concept, we shall discuss the integration of one such
tool, namely YASK~\cite{yask-main}, with Devito.

\subsection{Devito and seismic imaging}
Devito is a general purpose system, not restricted to specific PDEs, so it can
be used for any form of the wave equation.  Thus, unlike software specialized
in seismic exploration, like IWAVE~\cite{iwave} and
Madagascar~\cite{madagascar}, it suffers neither from the restriction to a
small set of wave equations and discretizations, nor from the lack of
portability and composability typical of a pure C/Fortran environment.

\subsection{Performance optimizations}
The Devito compiler can introduce three types of performance optimizations:
FLOP reduction, data locality, and parallelism. Typical FLOP reduction
transformations are common sub-expressions elimination, factorization, and code
motion. A thorough review is provided in~\cite{DSE-ref-glore}. Devito applies
all of these techniques (see Section~\ref{sec:DSE}). Particularly relevant for
stencil computation is the search for redundancies across consecutive loop
iterations~\cite{SIE-1,SIE-2,SIE-3}. This is at the core of the strategy
described in Section~\ref{sec:sie}, which essentially extends these ideas
with optimizations for data locality. Typical loop transformations for
parallelism and data locality~\cite{HPC-pearls} are also automatically
introduced by the Devito compiler (e.g., loop blocking, vectorization); more
details will be provided in Sections~\ref{sec:DLE} and~\ref{sec:YASK}.

\section{Specification of a finite-difference method with Devito}
\label{sec:api}
\input{symbolics}

\section{The Devito compiler}
\label{sec:compiler}
\begin{figure}
\begin{CenteredBox}
\includegraphics[scale=0.35]{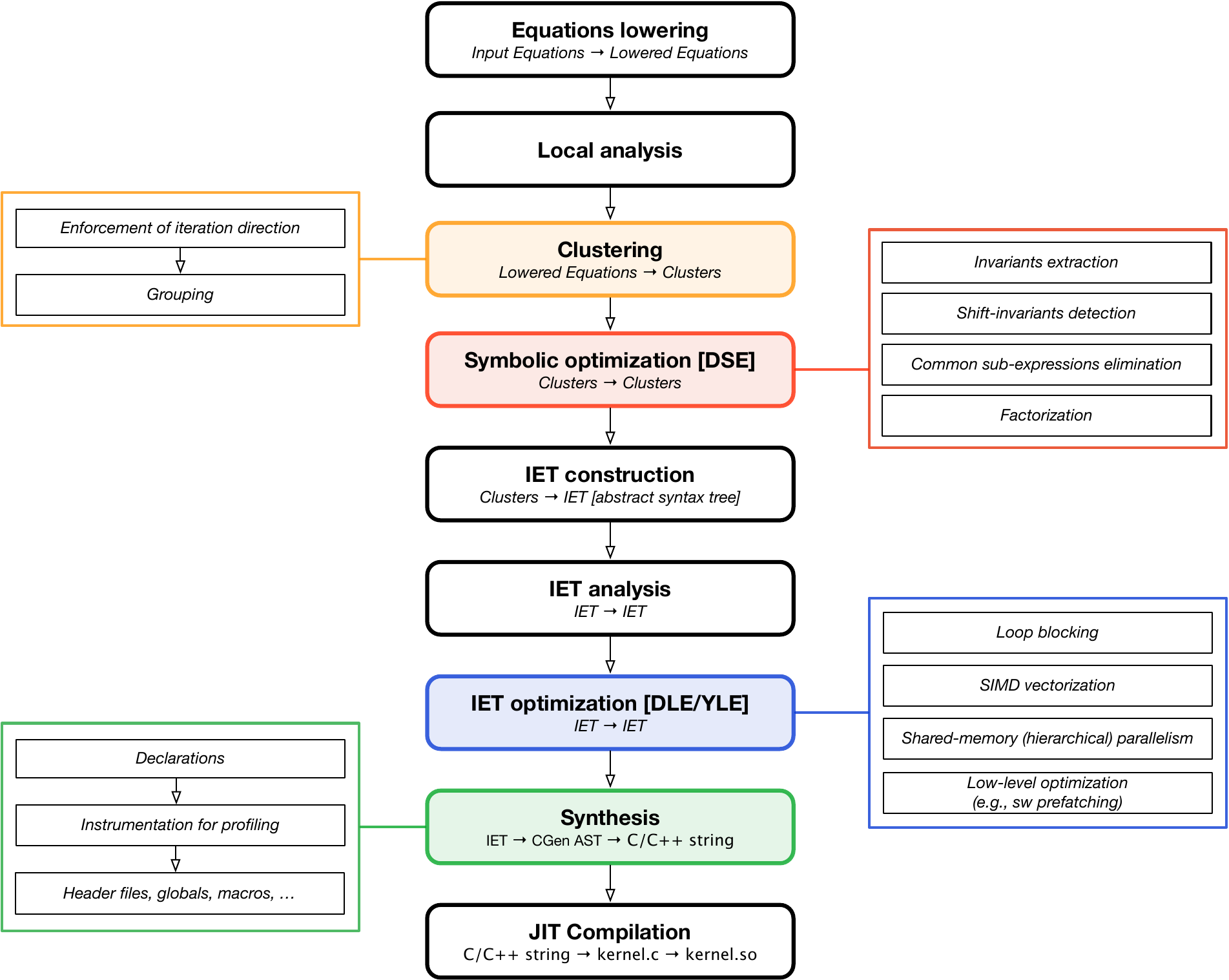}
\end{CenteredBox}
\caption{Compiler passes to lower symbolic equations into shared objects through an \Operator.}
\label{fig:devito-compiler}
\end{figure}

In Devito, an \Operator carries out three fundamental tasks:
generation of low-level code, JIT compilation, and
execution. The \Operator input consists of one or more symbolic
equations. In the generated code, these equations are scheduled within
loop nests of suitable depth and extent. The \Operator also accepts
substitution rules (to replace symbols with constant values) and
optimization levels for the Devito Symbolic Engine (DSE) and the
Devito Loop Engine (DLE). By default, all DSE and DLE optimizations
that are known to unconditionally improve performance are
automatically applied. The same \Operator may be reused with different
input data; JIT-compilation occurs only once, triggered by the first
execution.  Overall, this lowering process---from high-level
equations to dynamically compiled and executable code---consists of
multiple compiler passes, summarized in
Figure~\ref{fig:devito-compiler} and discussed in the following
sections (a minimal background in data dependence-analysis is
recommended; the unfamiliar reader may refer to a classic textbook
such as~\cite{dragonbook}).

\subsection{Equation lowering}
\label{sec:lowering}
In this pass, three main tasks are carried out: {\it indexification}, {\it
substitution}, and {\it domain-alignment}.
\begin{itemize}
\item As explained in Section~\ref{sec:api}, the input equations typically
    involve one or more indexed \Function{s}. The {\it indexification} consists
        of converting such objects into actual arrays. An array always keeps a
        reference to its originating \Function. For instance, all accesses to
        $u$ such as $u[t, x+1]$ and $u[t+1, x-2]$ would store a pointer to the
        same, user-defined \Function $u(t, x)$. This metadata is exploited
        throughout the various compilation passes.

\item During {\it substitution}, the user-provided substitution rules are
    applied. These may be given for any literal appearing in the input
        equations, such as the grid spacing symbols. Applying a substitution
        rule increases the chances of constant folding, but it makes the
        \Operator less generic. The values of symbols for which no substitution
        rule is available are provided at execution time.

\item The {\it domain-alignment} step shifts the array accesses deriving from
    \Function{s} having non-empty halo and padding regions. Thus, the array
        accesses become logically aligned to the equation's natural domain. For
        instance, given the usual \Function $u(t, x)$ having two points on each
        side of the $x$ halo region, the array accesses $u[t, x]$ and $u[t,
        x+2]$ are transformed, respectively, into $u[t, x+2]$ and $u[t, x+4]$.
        When $x=0$, therefore, the values $u[t, 2]$ and $u[t, 4]$ are fetched,
        representing the first and third points in the computational domain.
\end{itemize}

\subsection{Local analysis}
\label{sec:analysis}

The lowered equations are analyzed to collect information relevant for the
\Operator construction and execution. In this pass, an equation is inspected
``in isolation'', ignoring its relationship with the rest of the input. The
following metadata are retrieved and/or computed:
\begin{itemize}
\item input and output \Function{s};
\item \Dimension{s}, which are topologically ordered based on how they appear
    in the various array index functions; and
\item two notable \Space{s}: the iteration space, \ISpace, and the data space,
    \DSpace.
\end{itemize}

A \Space is a collection of points given by the product of $n$ compact
intervals on $\mathbb{Z}$. With the notation $d[o_m,o_M]$ we indicate the
compact interval $[d_m+o_m, d_M+o_M]$ over the \Dimension $d$, in which $d_m$
and $d_M$ are parameters (specialized only at runtime), while $o_m$ and $o_M$
are known integers. For instance, $[x[0, 0], y[-1, 1]]$ describes a rectangular
two-dimensional space over $x$ and $y$, whose points are given by the Cartesian
product $[x_m, x_M] \times [y_m - 1, y_M + 1]$. The \ISpace and \DSpace are two
special types of \Space. They usually span different sets of \Dimension{s}. A
\DSpace may have \Dimension{s} that do not appear in an \ISpace, in particular
those that are accessed only via integer indices.  Likewise, an \ISpace may
have \Dimension{s} that are not part of the \DSpace, such as a reduction axis.
Further, an \ISpace also carries, for each \Dimension, its iteration direction.

As an example, consider the equation {\it stencil} in Listing~\ref{lst:fd-3}.
Immediately we see that $\text{input} = [u, m]$, $\text{output} = [u]$,
$\Dimension{s} = [t, x]$. The compiler constructs the \ISpace $[t[0, 0]^{+},
x[0, 0]^{*}]$. The first entry $t[0, 0]^{+}$ indicates that, along $t$, the
equation should run between $t_m+0$ and $t_M+0$ (extremes included) in the {\it
forward} direction, as indicated by the symbol $+$. This is due to the fact
that there is a flow dependency in $t$, so only a unitary positive stepping
increment (i.e., $t = t + 1$) allows a correct propagation of information
across consecutive iterations. The only difference along $x$ is that the
iteration direction is now arbitrary, as indicated by $*$. The \DSpace is
$[t[0, 1], x[0, 0]]$; intuitively, the entry $t[0, 1]$ is used right before
running an \Operator to provide a default value for $t_M$---in particular,
$t_M$ will be set to the largest possible value that does not cause
out-of-domain accesses (i.e., out-of-bounds array accesses).

\subsection{Clustering}
\label{sec:clusterization}

A \Cluster is a sequence of equations having (i) same \ISpace, (ii) same
control flow (i.e., same \ConditionalDimension{s}), and (iii) no
dimension-carried ``true'' anti-dependencies among them.

As an example, consider again the setup in Section~\ref{sec:api}. The equation
{\it stencil} cannot be ``clusterized'' with the equations in the {\it
injection} list as their \ISpace{s} are different. On the other hand, the
equations in {\it injection} can be grouped together in the same \Cluster as
(i) they have same \ISpace $[t[0, 0]^{*}, p_q[0, 0]^{*}]$, (ii) same control
flow, and (iii) there are no true anti-dependencies among them (note that the
second equation in {\it injection} does write to $u[t+1, ...]$, but as
explained later this is in fact a reduction, that is a ``false''
anti-dependency).

\subsubsection{Iteration direction}
First, each equation is assigned a new \ISpace, based upon a {\it global}
analysis. Any of the iteration directions that had been marked as ``arbitrary''
($*$) during local analysis may now be enforced to {\it forward} ($+$) or {\it
backward} ($-$). This process exploits data dependence analysis.

For instance, consider the flow dependency between {\it stencil} and the {\it
injection} equations. If we want $u$ to be up-to-date when evaluating {\it
injection}, then we eventually need all equations to be scheduled sequentially
within the $t$ loop. For this, the \ISpace{s} of the {\it injection} equations
are specialized by enforcing the direction {\it forward} along the \Dimension
$t$. The new \ISpace is $[t[0, 0]^{+}, p_q[0, 0]^{*}]$.

Algorithm~\ref{algo:direction-derivation} illustrates how the enforcement of
iteration directions is achieved in general. Whenever a clash is detected
(i.e., two equations with \ISpace $[d[0, 0]^{+}, ...]$ and $[d[0, 0]^{-},
...]$), the original direction determined by the local analysis pass is kept
(lines~\ref{algo:direction-derivation:def1}
and~\ref{algo:direction-derivation:def2}), which will eventually lead to
generating different loops.

\begin{algorithm2e}
\caption{Clustering: enforcement of iteration directions (pseudocode).}\label{algo:direction-derivation}
\kwInput{A sequence of equations $\mathcal{E}$.}
\kwOutput{A sequence of equations $\mathcal{E}'$ with altered \ISpace.}
\Comment{Map each dimension to a set of expected iteration directions}
mapper $\gets$ {\sc detect$\_$flow$\_$directions}($\mathcal{E}$)\;
\For{e \In $\mathcal{E}$}{
    \For{dim, directions \In mapper}{
        \uIf{len(directions) == 1}{
            \Comment{No ambiguity}
            forced[dim] $\gets$ directions.pop()\;
        }\uElseIf{len(directions) == 2}{
            \Comment{No ambiguity as long as one of the two items is /Any/}
            \Try{
                directions.remove(Any)\;
                forced[dim] $\gets$ directions.pop()\;
            }\Except{
                forced[dim] $\gets$ e.directions[dim]\;\label{algo:direction-derivation:def1}
            }
        }\Else{
            forced[dim] $\gets$ e.directions[dim]\;\label{algo:direction-derivation:def2}
        }
    }
    $\mathcal{E}'$.append(e.$\_$rebuild(directions=forced))
}
\Return{$\mathcal{E'}$}
\end{algorithm2e}

\subsubsection{Grouping}
This step performs the actual clustering, checking \ISpace{s} and
anti-dependencies, as well as handling control flow. The procedure is shown in
Algorithm~\ref{algo:grouping}; some explanations follow.

\begin{algorithm2e}
\caption{Clustering: grouping expressions into \Cluster{s} (pseudocode)}\label{algo:grouping}
\kwInput{A sequence of equations $\mathcal{E}$.}
\kwOutput{A sequence of clusters $\mathcal{C}$.}
$\mathcal{C}$ $\gets$ ClusterGroup()\;
\For{e \In $\mathcal{E}$}{
    grouped $\gets$ \False\;
    \For{c \In reversed($\mathcal{C}$)}{
        anti, flow $\gets$ {\sc get$\_$dependencies}(c, e)\;
        \uIf{e.ispace == c.ispace \And anti.carried \Is empty}{\label{algo:grouping:ispace}
            c.add(e)\;
            grouped $\gets$ \True\;
            \Break\;
        }\uElseIf{anti.carried \Is \Not empty}{\label{algo:grouping:anti}
			c.atomics.update(anti.carried.cause)\;\label{algo:grouping:atomic}
			\Break\;
        }\uElseIf{flow.cause.intersection(c.atomics)}{\label{algo:grouping:flow}
            \Comment{cannot search across earlier clusters}
            \Break\;
        }
    }
    \If{\Not grouped}{
        $\mathcal{C}$.append(Cluster(e))\;
    }
}
$\mathcal{C}$ $\gets$ {\sc control$\_$flow}($\mathcal{C}$)\;
\Return{$\mathcal{C}$}
\end{algorithm2e}


\begin{itemize}

\item Robust data-dependence analysis, capable of tracking flow-, anti-, and
    output-dependencies at the level of array accesses, is necessary. In
        particular, it must be able to tell whether two generic {\it array
        accesses} induce a dependency or not.  The data-dependence analysis
        performed is conservative; that is, a dependency is always assumed when
        a test is inconclusive. Dependence testing is based on the standard
        Lamport test~\cite{dragonbook}. In Algorithm~\ref{algo:grouping}, 
        data-dependence analysis is carried out by the function {\sc
        get$\_$dependencies}.
        
\item If an anti-dependency is detected along a \Dimension $i$, then $i$ is
    marked as {\it atomic}---meaning that no further clustering can occur
        along $i$.  This information is also exploited by later \Operator
        passes (see Section~\ref{sec:scheduling}).

\item Reductions, and in particular increments, are treated specially. They
    represent a special form of anti-dependency, as they do not break
        clustering. {\sc get$\_$dependen\-ces} detects reductions and removes  
        them from the set of anti-dependencies.

\item Given the sequence of equations $[E_1, E_2, E_3]$, it is possible that
    $E_3$ can be grouped with $E_1$, but not with its immediate predecessor $E_2$
        (e.g., due to a different \ISpace). However, this can only happen when
        there are no flow or anti-dependen\-ces between $E_2$ and $E_3$; i.e.
        when the {\tt if} commands at lines~\ref{algo:grouping:anti}
        and~\ref{algo:grouping:flow} are not entered, thus allowing the search
        to proceed with the next equation. This optimization was
        originally motivated by gradient operators in seismic imaging kernels.

\item The routine {\sc control$\_$flow}, omitted for brevity, creates
    additional \Cluster{s} if one or more \ConditionalDimension{s} are
        encountered. These are tracked in a special \Cluster field, {\it
        guards}, as also required by later passes (see
        Section~\ref{sec:scheduling}).
\end{itemize}

\subsection{Symbolic optimization}
The DSE---Devito Symbolic Engine---is a macro-pass reducing the
\emph{arithmetic strength} of \Cluster{s} (e.g., their operation count). It
consists of a series of passes, ranging from standard common sub-expression
elimination (CSE) to more advanced rewrite procedures, applied individually to
each \Cluster. The DSE output is a new ordered sequence of \Cluster{s}: there
may be more or fewer \Cluster{s} than in the input, and both the overall number
of equations as well as the sequence of arithmetic operations might differ.
The DSE passes are discussed in Section~\ref{sec:DSE}. We remark
that the DSE only operates on \Cluster{s} (i.e., on collections of equations);
there is no concept of ``loop'' at this stage yet. However, by altering
\Cluster{s}, the DSE has an indirect impact on the final loop-nest structure.

\subsection{IET construction}
\label{sec:scheduling}
In this pass, the intermediate representation is lowered to an
Iteration/Expression Tree (IET). An IET is an abstract syntax tree in which
\Iteration{s} and \Expression{s}---two special node types---are the main
actors. Equations are wrapped within \Expression{s}, while \Iteration{s}
represent loops. Loop nests embedding such \Expression{s} are constructed by
suitably nesting \Iteration{s}. Each \Cluster is eventually placed in its own
loop (\Iteration) nest, although some (outer) loops may be shared by multiple
\Cluster{s}.

\begin{algorithm2e}
    \caption{An excerpt of the cluster scheduling algorithm, turning a list (of
    \Cluster{s}) into a tree (IET). Here, the fact that different \Cluster{s}
    may eventually share some outer \Iteration{s} is highlighted.}\label{algo:scheduling}
\kwInput{A sequence of \Cluster{s} $\mathcal{C}$.}
\kwOutput{An Iteration/Expression Tree.}
schedule $\gets$ list()\;
\For{c \In $\mathcal{C}$}{
    root $\gets$ None\;
    index $\gets$ 0\;
    \For{$i_0$, $i_1$ \In zip(c.ispace, schedule)}{\label{algo:scheduling:compare}
        \If{$i_0 \neq i_1$ \Or $i_0$.dimension \In c.atomics}{\label{algo:scheduling:atomic}
            \Break\;\label{algo:scheduling:break1}
        }
        root $\gets$ schedule[i1]\;
        index $\gets$ index + 1\;
        \If{$i_0$.dim \In c.guards}{
            \Break\;\label{algo:scheduling:break2}
        }
    }
    $\langle$build as many \Iteration{s} as \Dimension{s} in {\tt c.ispace[index:]} and nest them inside {\tt root}$\rangle$\; \label{algo:scheduling:create}
    $\langle$update {\tt schedule}$\rangle$\;
    $\langle$...$\rangle$
}
\end{algorithm2e}

Consider again our running acoustic wave-equation example. There are three
\Cluster{s} in total: $C_1$ for {\it stencil}, $C_2$ for {\it save}, and $C_3$
for the equations in {\it injection}. We use Algorithm~\ref{algo:scheduling}---an
excerpt of the actual cluster scheduling algorithm---to explain how this
sequence of \Cluster{s} is turned into an IET. Initially, the {\it schedule}
list is empty, so when $C_1$ is handled two nested \Iteration{s} are created
(line~\ref{algo:scheduling:create}), respectively for the \Dimension{s} $t$ and
$x$. Subsequently, $C_2$'s \ISpace and the current {\it schedule} are compared
(line~\ref{algo:scheduling:compare}). It turns out that $t$ appears among
$C_2$'s guards, hence the for loop is exited at
line~\ref{algo:scheduling:break2} without inspecting the second and last
iteration. Thus, $index = 1$, and the previously built \Iteration over $t$ is
reused. Finally, when processing $C_3$, the for loop is exited at the second
iteration due to line~\ref{algo:scheduling:atomic}, since $p_q \neq x$.
Again, the $t$ \Iteration is reused, while a new \Iteration is constructed for
the \Dimension $p_q$. Eventually, the constructed IET is as in
Listing~\ref{lst:cluster-scheduling-example-3}.

\begin{listing}
\begin{lstlisting}
for t = t_m to t_M:
 |-- for x = x_m to x_M:
 |    |-- <Eq(u[t+1,x], ...)>
 |
 |-- if t % 4 == 0
 |    |-- for x = x_m to x_M:
 |         |-- <Eq(us[t/4, x], ...)>
 |
 |-- for p_q = p_q_m to p_q_M:
      |-- <Eq(u[t+1,f(p_q)], ...)>
      |-- <Eq(u[t+1,g(p_q)], ...)>

\end{lstlisting}
\caption{Graphical representation of the IET produced by the cluster scheduling
    algorithm for the running example.}
\label{lst:cluster-scheduling-example-3}
\end{listing}

\subsection{IET analysis}
The newly constructed IET is analyzed to determine \Iteration properties such
as {\tt sequential}, {\tt parallel}, and {\tt vectorizable}, which are
then attached to the relevant nodes in the IET. These properties are used
for loop optimization, but only by a later pass (see
Section~\ref{sec:loop-opt}). To determine whether an \Iteration is {\tt
parallel} or {\tt sequential}, a fundamental result from compiler theory is
used---the $i$-th \Iteration in a nest comprising $n$ \Iteration{s} is
parallel if for all dependencies $D$, expressed as distance vectors $D = (d_0,
..., d_{n-1})$, either $(d_1, ..., d_{i-1}) > 0$ or $(d_1, ..., d_i) =
0$ ~\cite{dragonbook}. 

\subsection{IET optimization}
\label{sec:loop-opt}
This macro-pass transforms the IET for performance optimization. Apart from
runtime performance, this pass also optimizes for rapid JIT compilation with
the underlying C compiler. A number of loop optimizations are introduced,
including loop blocking, minimization of remainder loops, SIMD vectorization,
shared-memory (hierarchical) parallelism via OpenMP, software prefetching.
These will be detailed in Section~\ref{sec:optimizations}. A {\it backend} (see
Section~\ref{sec:backends}) might provide its own loop optimization engine.

\subsection{Synthesis, dynamic compilation, and execution}
\label{sec:synthesis}
Finally, the IET adds variable declarations and header files, as well as
instrumentation for performance profiling, in particular, to collect execution
times of specific code regions. Declarations are injected into the
IET, ensuring they appear as close as possible to the scope in which the
relative variables are used, while honoring the OpenMP semantics of private and
shared variables. To generate C code, a suitable tree visitor inspects the IET
and incrementally builds a \CGen tree~\cite{cgen}, which is ultimately
translated into a string and written to a file. Such files are stored in a
software cache of Devito-generated \Operator{s}, JIT-compiled into a shared
object, and eventually loaded into the \Python environment. The compiled code
has a default entry point (a special function), which is called directly from
\Python at \Operator application time.

\subsection{Operator specialization through backends}
\label{sec:backends}
In Devito, a \emph{backend} is a mechanism to specialize data types as well as
\Operator passes, while preserving software modularity (inspired by~\cite{pyop2}).

One of the main objectives of the backend infrastructure is promoting software
composability. As explained in Section~\ref{sec:related-work}, there exist a
significant number of interesting tools for stencil optimization, which we may
want to integrate with Devito.  For example, one of the future goals is to
support GPUs, and this might be achieved by writing a new backend implementing
the interface between Devito and third-party software specialized for this
particular architecture. 

Currently, two backends exist:
\begin{description}
\item[\core] the default backend, which relies on the DLE for loop
    optimization. 
\item[\yask] an alternative backend using the YASK stencil compiler to generate
    optimized C++ code for \Intel\xspace \Xeon and \XeonPhi
        architectures~\cite{yask-main}. Devito transforms the IET into a
        format suitable for YASK, and uses its API for data management,
        JIT-compilation, and execution. Loop optimization is performed by YASK
        through the YASK Loop Engine (YLE).
\end{description}
The \core and \yask backends share the compilation pipeline in
Figure~\ref{fig:devito-compiler} until the loop optimization stage.

\section{Automated performance optimizations}
\label{sec:optimizations}
As discussed in Section~\ref{sec:compiler}, Devito performs symbolic
optimizations to reduce the arithmetic strength of the expressions, as well as
loop transformations for data locality and parallelism. The former are
implemented as a series of compiler passes in the DSE, while for the latter
there currently are two alternatives, namely the DLE and the YLE (depending on
the chosen execution backend).

Devito abstracts away the single optimizations passes by providing users with a
certain number of optimization levels, called ``modes``, which trigger
pre-established sequences of optimizations---analogous to what general-purpose
compilers do with, for example, {\tt -O2} and {\tt -O3}.  In
Sections~\ref{sec:DSE},~\ref{sec:DLE}, and~\ref{sec:YASK} we describe the
individual passes provided by the DSE, DLE, and YLE respectively, while in
Section~\ref{sec:performance:cs} we explain how these are composed into
modes.

\subsection{DSE---Devito Symbolic Engine}
\label{sec:DSE}
The DSE passes attempt to reduce the arithmetic strength of the expressions
through FLOP-reducing transformations~\cite{DSE-ref-glore}. They are
illustrated in Listings~\ref{lst:dse-init}-\ref{lst:dse-end}, which derive from
the running example used throughout the article. A detailed description
follows.

\begin{itemize}
\item {\bf Common sub-expression elimination (CSE).} Two implementations are
    available: one based upon \Sympy 's {\tt cse} routine and one built on
        top of more basic \Sympy routines, such as {\tt xreplace}. The former
        is more powerful, being aware of key arithmetic properties such as
        associativity; hence it can discover more redundancies.  The latter is
        simpler, but avoids a few critical issues: (i) it has a much quicker
        turnaround time; (ii) it does not capture integer index expressions
        (for increased quality of the generated code); and (iii) it tries not
        to break factorization opportunities. A generalized common
        sub-expressions elimination routine retaining the features and avoiding
        the drawbacks of both implementations is still under development. By
        default, the latter implementation is used when the CSE pass is
        selected.

\begin{listing}
\begin{lstlisting}[mathescape=true]
>>> 9.0*dt*dt*u[t, x + 1] - 18.0*dt*dt*u[t][x + 2] + 9.0*dt*dt*u[t, x + 3]
temp0 = dt*dt
9.0*temp0*u[t, x + 1] - 18.0*temp0*u[t][x + 2] + 9.0*temp0*u[t, x + 3]
\end{lstlisting}
\caption{An example of common sub-expressions elimination.}
\label{lst:dse-init}
\end{listing}

\item {\bf Factorization.} This pass visits each expression tree and tries to
    factorize FD weights. Factorization is applied without altering the
        expression structure (e.g., without expanding products) and without
        performing any heuristic search across groups of expressions. This
        choice is based on the observation that a more aggressive approach is
        only rarely helpful (never in the test cases in
        Section~\ref{sec:performance}), while the increase in symbolic
        processing time could otherwise be significant. The implementation
        exploits the \Sympy\xspace\xspace{\tt collect} routine. However, while
        {\tt collect} only searches for common factors across the immediate
        children of a single node, the DSE implementation recursively applies
        {\tt collect} to each {\tt Add} node (i.e., an addition) in the
        expression tree, until the leaves are reached.

\begin{listing}
\begin{lstlisting}[mathescape=true]
>>> 9.0*temp0*u[t, x + 1] - 18.0*temp0*u[t][x + 2] + 9.0*temp0*u[t, x + 3]
9.0*temp0*(u[t, x + 1] + u[t, x + 3]) - 18.0*temp0*u[t][x + 2]
\end{lstlisting}
\caption{An example of FD weights factorization.}
\end{listing}

\item {\bf Extraction.} The name stems from the fact that sub-expressions
    matching a certain condition are pulled out of a larger expression, and
        their values are stored into suitable scalar or tensor temporaries.
        For example, a condition could be ``{\it extract all time-varying
        sub-expressions whose operation count is larger than a given
        threshold}''. A tensor temporary may be preferred over a scalar
        temporary if the intention is to let the {\it IET construction} pass
        (see Section~\ref{sec:scheduling}) place the pulled sub-expressions
        within an outer loop nest. Obviously, this comes at the price
        of additional storage. This peculiar effect---trading operations for
        memory---will be thoroughly analyzed in Sections~\ref{sec:sie}
        and~\ref{sec:performance}.

\begin{listing}
\begin{lstlisting}[mathescape=true]
>>> 9.0*temp0*(u[t, x + 1] + u[t, x + 3]) - 18.0*temp0*u[t][x + 2]
temp1[x] = u[t, x + 1] + u[t, x + 3]
9.0*temp0*temp1[x] - 18.0*temp0*u[t][x + 2]
\end{lstlisting}
\caption{An example of time-varying sub-expressions extraction. Only
    sub-expressions performing at least one floating-point operation are
    extracted.}
\end{listing}

\item {\bf Detection of shift-invariants.} In essence, a shift-invariant is a
    sub-expression that is redundantly computed at multiple iteration points.
    Because of its key role in the Shift-invariants Elimination algorithm, the
    explanation of how shift-invariants are detected is postponed until
    Section~\ref{sec:sie}.

\begin{listing}
\begin{lstlisting}[mathescape=true]
>>> 9.0*temp0*u[t, x + 1] - 18.0*temp0*u[t][x + 2] + 9.0*temp0*u[t, x + 3]
temp[x] = 9.0*temp0*u[t, x]
temp[x + 1] - 18.0*temp0*u[t][x + 2] + temp[x + 3] 
\end{lstlisting}
\caption{An example of shift-invariant detection. The shift-invariant
    $9.0*temp0*u[t, x]$ is assigned to the vector temporary $temp[x]$ so that
    it can be used in place of the two sub-expressions $9.0*temp0*u[t, x + 1]$
    and $9.0*temp0*u[t, x + 3]$.}
\label{lst:dse-end}
\end{listing}

\end{itemize}

\subsection{DLE---Devito Loop Engine}
\label{sec:DLE}
The DLE transforms the IET via classic loop optimizations for parallelism and
data locality~\cite{HPC-pearls}. These are summarized below.

\begin{itemize}
\item {\bf SIMD Vectorization.} Implemented by enforcing compiler
    auto-vectorization via special {\tt pragma}s from the OpenMP 4.0 language.
        With this approach, the DLE aims to be performance-portable across
        different architectures. However, this strategy causes a significant
        fraction of vector loads/stores to be unaligned to cache boundaries,
        due to the stencil offsets. As we shall see, this is a primary cause of
        performance loss.

\item {\bf Loop Blocking.} Also known as ``tiling'', this technique
        implemented by replacing \Iteration trees in the IET.
        The current implementation only supports blocking over fully-parallel
        \Iteration{s}.  Blocking over dimensions characterized by flow- or
        anti-dependencies, such as the time dimension in typical explicit finite
        difference schemes, is instead work in progress (this would require a
        preliminary pass known as loop skewing; see
        Section~\ref{sec:further-work} for more details). On the other hand, a
        feature of the present implementation is the capability of blocking
        across particular {\it sequences} of loop nests. This is exploited by
        the Shift-invariants Elimination algorithm, as shown in
        Section~\ref{sec:sie:blocking}. To determine an optimal block shape,
        an \Operator resorts to empirical auto-tuning. 

\item {\bf Parallelism.} Shared-memory parallelism is introduced by decorating
    \Iteration{s} with suitable OpenMP {\tt pragma}s. The OpenMP {\tt static}
        scheduling is used. Normally, only the outermost fully-parallel
        \Iteration is annotated with the parallel {\tt pragma}. However, heuristically
        nested fully-parallel \Iteration{s} are {\tt collapsed} if the
        core count is greater than a certain threshold. This pass also ensures
        that all array temporaries allocated in the scope of the parallel
        \Iteration are declared as {\tt private} and that storage is allocated
        where appropriate (stack, heap). 
\end{itemize}

Summarizing, the DLE applies a sequence of typical stencil optimizations,
aiming to reach a minimum level of performance across different architectures.
As we shall see, the effectiveness of this approach, based on simple
transformations, deteriorates on architectures strongly conceived for
hierarchical parallelism. This is one of the main reasons behind the
development of the \yask backend (see Section~\ref{sec:backends}), described in
the following section.

\subsection{YLE---YASK Loop Engine}
\label{sec:YASK}
\input{yask}

\section{The Shift-invariants Elimination Algorithm}
\label{sec:sie}
Shift-invariants, or ``cross-iteration redundancies'' (informally introduced in
Section~\ref{sec:DSE}), in FD operators depend on the differential operators
used in the PDE(s) and the chosen discretization scheme. From a performance
viewpoint, the presence of shift-invariants is a non-issue as long as the operator is
memory-bound, while it becomes relevant in kernels with a high arithmetic
intensity. In Devito, the Shift-invariants Elimination (SIE)
algorithm attempts to remove shift-invariants with the goal of reducing the operation
count. As shown in Section~\ref{sec:performance}, the SIE algorithm has
considerable impact in seismic imaging kernels. The algorithm is implemented
through the orchestration of multiple DSE and DLE/YLE passes, namely {\it
extraction of candidate expressions (DSE)}, {\it detection of shift-invariants (DSE)},
{\it loop blocking (DLE/YLE)}.

\subsection{Extraction of candidate expressions}
\label{sec:sie:extraction}
The criteria for extraction of candidate sub-expressions are:
\begin{itemize}
\item Any {\it maximal time-invariant} whose operation count is greater than
    $Thr_0=10$ (floating point arithmetic only). The term ``maximal''  means
        that the expression is not embedded within a larger time-invariant. The
        default value $Thr_0=10$, determined empirically, provides systematic
        performance improvements in a series of seismic imaging kernels.
        Transcendental functions are given a weight in the order of tens of
        operations, again determined empirically.

\item Any {\it maximal time-varying} whose operation count is greater than
    $Thr_1=10$. Such expressions often lead to shift-invariants, since they typically
        result from taking spatial and time derivatives on \TimeFunction{s}. In
        particular, cross-derivatives are a major cause of shift-invariants.
\end{itemize}

This pass leverages the {\it extraction} routine described in
Section~\ref{sec:DSE}.

\subsection{Detection of shift-invariants}
\label{sec:sie:ada}
To define the concept of shift-invariant expressions, we first need to formalize the
notion of {\it shifted operands}. Here, an operand is regarded as the
arithmetic product of a scalar value (or ``coefficient'') and one or more
indexed objects. An indexed object is characterized by a label (i.e., its
name), a vector of $n$ dimensions, and a vector of $n$ displacements (one for
each dimension). We say that an operand $o_1$ is shifted with respect to an
operand $o_0$ if $o_0$ and $o_1$ have same coefficient, label, and dimensions,
and if their displacement vectors are such that one is the translation of the
other (in the classic geometric sense).  For example, the operand $2*u[x,y,z]$
is shifted with respect to the operand $2*u[x+1,y+2,z+3]$ since they have
same coefficient ($2$), label ($u$), and dimensions ($[x,y,z]$), while the
displacement vectors $[0,0,0]$ and $[1,2,3]$ are expressible by means of a
translation.

Now consider two expressions $e_0$ and $e_1$ in fully-expanded form (i.e., a
non-nested sum-of-operands). We say that $e_0$ is shifted with respect to $e_1$
if the following conditions hold:
\begin{itemize}
\item the operands in $e_0$ ($e_1$) are shifted with respect to the operands in
    $e_1$ ($e_0$);
\item the same arithmetic operators are applied to the involved operands.
\end{itemize}
For example, consider $e = u[x] + v[x]$, having two operands $u[x]$ and $v[x]$;
then:
\begin{itemize}
\item {\bf u[x-1] + v[y-1]} is {\it not} shifted with respect to $e$, due to a
    different dimension vector.
\item {\bf u[x] + w[x]} is {\it not} shifted with respect to $e$, due to a different label.
\item {\bf u[x+2] + v[x]} is {\it not} shifted with respect to $e$, since it
    cannot be expressed as a translation of $e$.
\item {\bf u[x+2] + v[x+2]} is shifted with respect to $e$, as it can be
    expressed through the translation $T(x) = x + \mathbf{2}$.
\end{itemize}

The relation ``$e_0$ is shifted with respect to $e_1$'' is an equivalence relation, as it
is at the same time reflexive, symmetric, and transitive. Thanks to these
properties, the turnaround times for detecting shift-invariants are extremely
quick (less than 2 seconds running on an \Intel\xspace \Xeon E5-2620 v4 for the
challenging \tti test case with \so{=}16, described in
Section~\ref{sec:performance:problems}), despite the $O(n^2)$ computational
complexity (with $n$ representing the number of candidate expressions, see
Section~\ref{sec:sie:extraction}).

Algorithm~\ref{algo:shift-invariants} highlights the fundamental steps
of shift-invariants detection. In the worst case scenario, all pairs of
candidate expressions are compared by applying the shift-invariant definition given
above. Aggressive pruning, however, is applied to minimize the cost of the
search. The algorithm uses some auxiliary functions: (i) {\sc
calculate$\_$displacements} returns a mapper associating, to each candidate,
its displacement vectors (one for each indexed object); (ii) {\sc
compare$\_$ops}($e_1, e_2$) evaluates to true if $e_1$ and $e_2$ perform the
same operations on the same operands; (iii) {\sc is$\_$translated}($d_1, d_2$)
evaluates to true if the displacement vectors in $d_2$ are pairwise-shifted
with respect to the vectors in $d_1$ by the same factor. Together, (ii) and
(iii) are used to establish whether two expressions are shifted
(line~\ref{algo:shift-invariants:compareto}). From an implementation point of view,
these functions exploit key \Sympy expression properties (e.g., immutability,
deterministic ordering of operands) and operators (e.g., for structural
equality testing), so they eventually result rather simple.

Eventually, $m$ sets of shift-invariants are determined. For each
of these sets $G_0, ..., G_{m-1}$, a \emph{pivot}---a special shift-invariant
-- is constructed. This is the key for operation count reduction:
the pivot $p_i$ of $G_i = \lbrace e_0, ..., e_{k-1}\rbrace$ will be used in
place of $e_0, ..., e_{k-1}$ (thus obtaining a reduction proportional to $k$).
A simple example is illustrated in Listing~\ref{lst:dse-end}.

\begin{algorithm2e}
\caption{Detection of shift-invariants (pseudocode).}\label{algo:shift-invariants}
\kwInput{A sequence of expressions $\mathcal{E}$.}
\kwOutput{A sequence of shift-invariant objects $\mathcal{A}$.}
    displacements $\gets$ {\sc calculate$\_$displacements}($\mathcal{E}$)\;
$\mathcal{A}$ $\gets$ list()\;
unseen $\gets$ list($\mathcal{E}$)\;
\While{unseen \Is \Not empty}{
    top $\gets$ unseen.pop()\;
    G = ShiftInvariant(top)\;
    \For{e \In unseen}{
        \If{{\sc compare$\_$ops}(top, e) \And {\sc is$\_$translated}(displacements[top], displacements[e])}{ \label{algo:shift-invariants:compareto}
            G.append(e)\;
            unseen.remove(e)\;
        }
    }
    $\mathcal{A}$.append(G)
}
\Return{$\mathcal{A}$}
\end{algorithm2e}

Several optimizations for data locality, not shown in
Algorithm~\ref{algo:shift-invariants}, are also applied. The interested reader may refer
to the documentation and the examples of \DevitoVer for more details; below, we
only mention the underlying ideas.

\begin{itemize}
\item The pivot of $G_i$ is {\it constructed}, rather than selected out of
    $e_0, ..., e_{k-1}$, so that it could coexist with as many other pivots as
        possible within the same \Cluster. For example, consider again
        Listing~\ref{lst:dse-end}: there are infinite possible pivots {\tt
        temp[x + s] = 9.0*temp0*u[t, x + s]}, and the one with $s=0$ is chosen.
        However, this choice is not random. The pivots are chosen based on
        a global optimization strategy, which takes into account all of the $m$
        sets of shift-invariants. The objective function consists of choosing $s$
        so that multiple pivots will have identical \ISpace, and thus be scheduled
        to the same \Cluster (and, eventually, to the same loop nest).

\item Conservatively, the chosen pivots are assigned to array variables. A
    second optimization pass, called {\it index bumping and array contraction}
        in \DevitoVer, attempts to turn these arrays into scalar variables,
        thus reducing memory consumption. This pass is based on data-dependence
        analysis, which essentially checks whether a given pivot is required
        only within its \Cluster or by later \Cluster{s} as well. In the former
        case, the optimization is applied.
\end{itemize}

\subsection{Loop blocking for working-set minimization}
\label{sec:sie:blocking}
In essence, the SIE algorithm trades operation for memory---the (array)
temporaries to store the shift-invariants. From a run-time performance viewpoint, this
is convenient only in arithmetic-intensive kernels. Unsurprisingly, we
observed that storing temporary arrays spanning the entire grid rarely provides
benefits (e.g., only when the operation count reductions are exceptionally
high). We then considered the following options.

\begin{enumerate}
\item Capturing redundancies arising along the innermost dimension only. Thus,
    only scalar temporaries would be necessary. This approach presents three
        main issues, however: (i) only a small percentage of all redundancies
        are captured; (ii) the implementation is non-trivial, due to the need
        for circular buffers in the generated code; (iii) SIMD vectorization is
        affected, since inner loop iterations are practically serialised.
        Some previous articles followed this path~\cite{SIE-1,SIE-2}.

\item A generalization of the previous approach: using both scalar and array
    temporaries, without searching for redundancies across the outermost
        loop(s). This mitigates issue (i), although the memory pressure is
        still severely affected. Issue (iii) is also unsolved.  This strategy
        was discussed in~\cite{SIE-3}.

\item Using loop blocking. Redundancies are sought and captured along all
    available dimensions, although they are now assigned to array temporaries
        whose size is a function of the block shape. A first loop nest produces
        the array temporaries, while a subsequent loop nest consumes them, to
        compute the actual output values. The block shape should be chosen so
        that writes and reads to the temporary arrays do not cause high latency
        accesses to the DRAM. An illustrative example is shown in
        Listing~\ref{lst:sie}.
\end{enumerate}

The SIE algorithm uses the third approach, based on cross-loop-nest blocking.
This pass is carried out by the DLE, which can introduce blocking over
sequences of loops (see Section~\ref{sec:DLE}).

\begin{listing}
\begin{lstlisting}[escapechar=|]
for t = t_m to t_M:
  for xb = x_m to x_M, xb += blocksize:|\label{lst:sie:block}|
    for x = xb to xb + blocksize + 3, x += 1|\label{lst:sie:l0}|
      temp[x] = 9.0*temp0*u[t, x]
    for x = xb to xb + blocksize; x += 1:|\label{lst:sie:l1}|
      u[t+1,x,y] = ... + temp[x + 1] - 18.0*temp0*u[t][x + 2] + temp[x + 3] + ...
\end{lstlisting}
\caption{The loop nest produced by the SIE algorithm for the example in
    Listing~\ref{lst:dse-end}. Note that the block loop
    (line~\ref{lst:sie:block}) wraps both the producer
    (line~\ref{lst:sie:l0}) and consumer (line~\ref{lst:sie:l1}) loops.
    For clarity, unnecessary information is omitted.}
\label{lst:sie}
\end{listing}

\section{Performance evaluation}
\label{sec:performance}
We outline in Section~\ref{sec:performance:cs} the compiler setup, computer
architectures, and measurement procedure that we used for our performance experiments.
Following that, we outline the physical model and numerical setup that define
the problem being solved in Section~\ref{sec:performance:problems}.
This leads to performance results, presented in
Sections~\ref{sec:performance:isotropic} and~\ref{sec:performance:tti}.

\subsection{Compiler and system setup}
\label{sec:performance:cs}
We analyze the performance of generated code using enriched roofline plots. Since
the DSE transformations may alter the operation count by allocating extra
memory, only by looking at GFlops/s performance and runtime jointly can
a quality measure of code syntheses be derived. 

For the roofline plots, Stream TRIAD was used to determine the attainable
memory bandwidth of the node. Two peaks for the maximum floating-point
performance are shown: the ideal peak, calculated as
\begin{align*}
\# \text{[cores]} \cdot \# \text{[avx units]} \cdot \# \text{[vector lanes]} \cdot \# \text{[FMA ports]} \cdot \text{[ISA base frequency]}
\end{align*}
and a more realistic one, given by the LINPACK benchmark. The reported runtimes
are the minimum of three runs (the variance was negligible). The model used to 
calculate the operational intensity assumes that the time-invariant
\Function{s} are reloaded at each time iteration. This is a more realistic
setting than a ``compulsory-traffic-only'' model (i.e., an infinite cache). 

We had exclusive access to two architectures: an \Skylake and an \KNL, which
will be referred to as \skl and \knl. Thread pinning was enabled with the
program {\tt numactl}. The \Intel compiler {\tt icc version 18.0} was used to compile
the generated code. The experiments were run with \DevitoVer~\cite{devito31}.
The experimentation framework with instructions for reproducibility is
available at~\cite{devito-performance}. All floating point operations are
performed in single precision, which is typical for seismic imaging
applications.

Any arbitrary sequence of DSE and DLE/YLE transformations is applicable to an
\Operator. Devito, provides three preset optimization sequences, or
``modes'', which vary in aggressiveness and affect code generation in three
major ways:
\begin{itemize}
\item the time required by the Devito compiler to generate the code,
\item the potential reduction in operation count, and
\item the potential amount of additional memory that might be allocated to
    store (scalar, tensor) temporaries.
\end{itemize}
A more aggressive mode might obtain a better operation count reduction than a
non-aggressive one, although this does not necessarily imply a better time to
solution as the memory pressure might also increase. The three optimization
modes---\optbasic, \optadv, and \optagg---apply the same sequence of DLE/YLE
transformations, which includes OpenMP parallelism, SIMD vectorization, and
loop blocking. However, they vary in the number, type, and order of DSE
transformations. In particular,
\begin{description}
\item[\optbasic] enables: common sub-expressions elimination.
\item[\optadv] enables: factorization; extraction of time-invariant
    shift-invariants; detection of shift-invariants; all \optbasic passes.
\item[\optagg] enables: extraction of time-varying shift-invariants; all
    \optadv passes.
\end{description}
Thus, \optagg triggers the full-fledged SIE algorithm, while \optadv uses only
a relaxed version (based on time invariants).
All runs used loop tiling with a block shape that was determined individually for
each case using auto-tuning. The auto-tuning phase, however, was not included in
the measured experiment runtime. Likewise, the code generation phase is not
included in the reported runtime.

\subsection{Test case setup}
\label{sec:performance:problems}
In the following sections, we benchmark the performance of operators modeling
the propagation of acoustic waves in two different models: isotropic and Tilted
Transverse Isotropy (TTI, \cite{zhang-tti}), henceforth \isotropic and \tti,
respectively. These operators were chosen for their relevance in seismic
imaging techniques~\cite{zhang-tti}.

\begin{figure}[htpb]
\begin{center}
\begin{tabulary}{\textwidth}{ C C C }
\includegraphics[width=3.7cm,keepaspectratio]{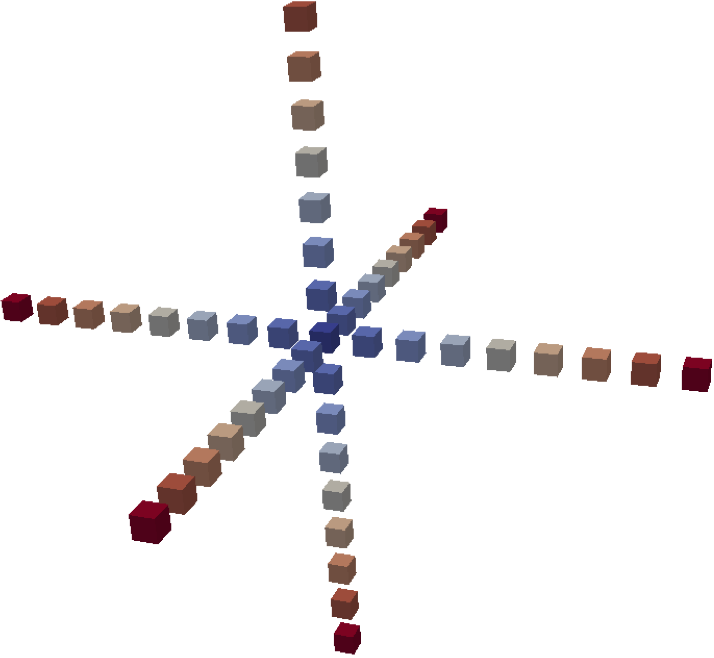} &
    ~~~~~ &
\includegraphics[width=3.7cm,keepaspectratio]{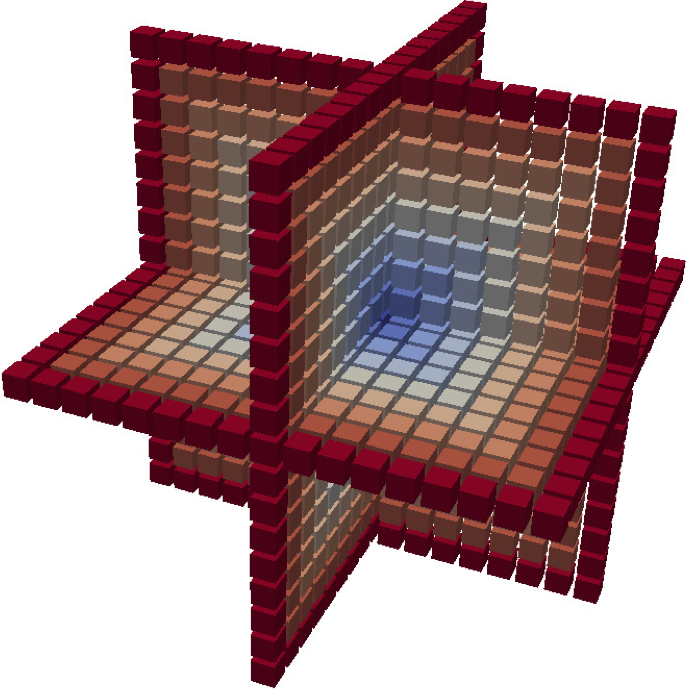}
\end{tabulary}
\end{center}
\caption{Stencils of the acoustic Laplacian for the \isotropic (left) and
\tti (right) wave equations and space-order of 16. The anisotropic Laplacian
corresponds to a spatially rotated version of the isotropic Laplacian. The
color indicates the distance from the central coefficient.}
\label{fig:stencil}
\end{figure}

Acoustic \isotropic modeling is the most commonly used technique for seismic
inverse problems, due to the simplicity of its implementation, as well as the
comparatively low computational cost in terms of FLOPs. The \tti wave equation
provides a more realistic simulation of wave propagation and accounts for local
directional dependency of the wave speed, but comes with increased
computational cost and mathematical complexity. For our numerical tests, we use
the \tti wave equation as defined in \cite{zhang-tti}. The full specification
of the equation as well as the finite difference schemes and its implementation
using Devito are provided in~\cite{devito-gmd, louboutin2016ppf}. Essentially,
the \tti wave equation consists of two coupled acoustic wave equations, in
which the Laplacians are constructed from spatially rotated first derivative
operators.  As indicated by Figure~\ref{fig:stencil}, these spatially rotated
Laplacians have a significantly larger number of stencil coefficients in
comparison to its isotropic equivalent which comes with an increased
operational intensity.

The \tti and \isotropic equations are discretized with second order in time and
varying space orders of 4, 8, 12 and 16. For both test cases, we use zero
initial conditions, Dirichlet boundary conditions and absorbing boundaries with
a 10 point mask (Section~\ref{sec:api:domain-halo}). The waves are excited by
injecting a time-dependent, but spatially-localized seismic source wavelet into
the subsurface model, using Devito's sparse point interpolation and injection
as described in Section~\ref{sec:api:symbolics}. We carry out performance
measurements for two velocity models of $512^3$ and $768^3$ grid points with a
grid spacing of 20 m. Wave propagation is modeled for 1000 ms, resulting in 327
time steps for \isotropic, and 415 time steps for \tti. The time-stepping
interval is chosen according to the Courant-Friedrichs-Lewy (CFL) condition
\cite{Courant1967}, which guarantees stability of the explicit time-marching
scheme and is determined by the highest velocity of the subsurface model and
the grid spacing.

\subsection{Performance: acoustic wave in isotropic model}
\label{sec:performance:isotropic}
This section illustrates the performance of \isotropic with the \core and
\yask backends. To simplify the exposition, we show results for the DSE in
\optadv mode only; the \optagg has no impact on \isotropic, due to the
memory-bound nature of the code~\cite{louboutin2016ppf}.

\begin{figure}[htpb]
\centering
\begin{subfigure}[b]{0.5\textwidth}
\includegraphics[width=\textwidth]{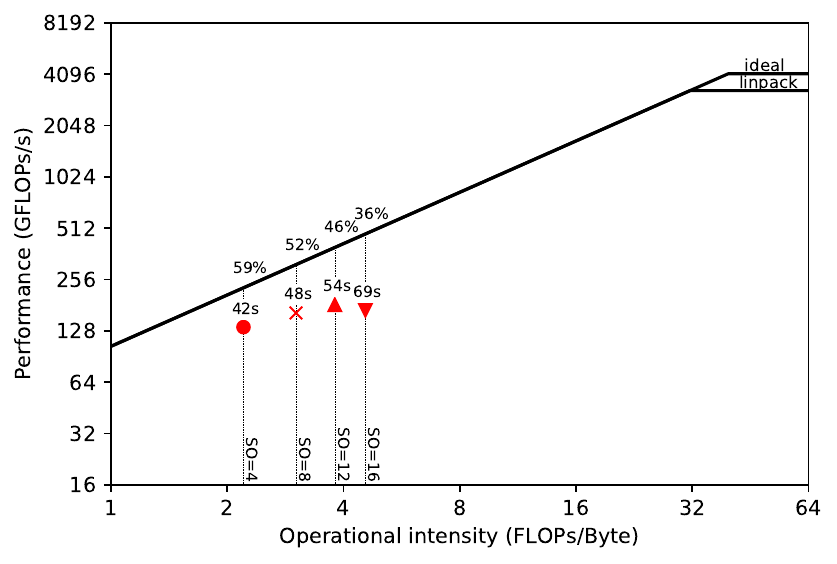}
\caption{\skl, \core, $768^3$ grid points.}\label{fig:acoustic:core-skl}
\end{subfigure}%
\begin{subfigure}[b]{0.5\textwidth}
\includegraphics[width=\textwidth]{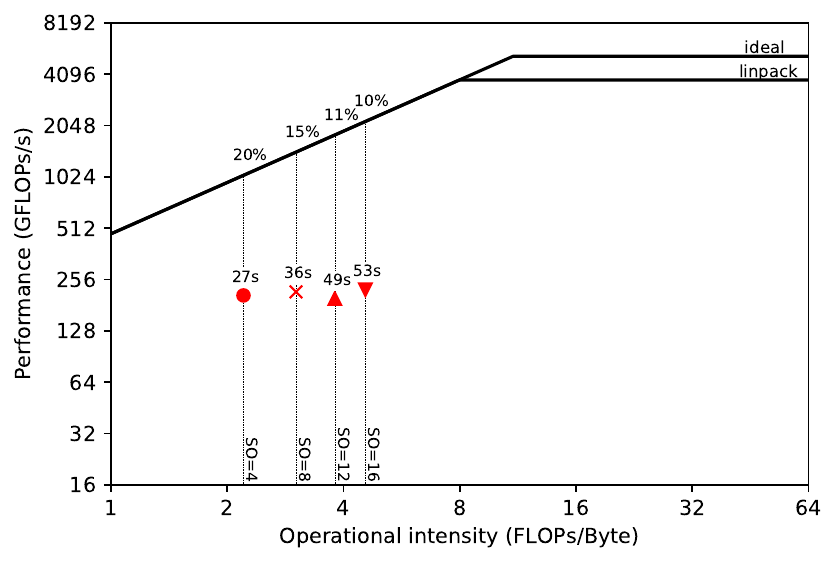}
\caption{\knl, \core, $768^3$ grid points.}\label{fig:acoustic:core-knl}
\end{subfigure}%
~\\
\begin{subfigure}[b]{0.5\textwidth}
\includegraphics[width=\textwidth]{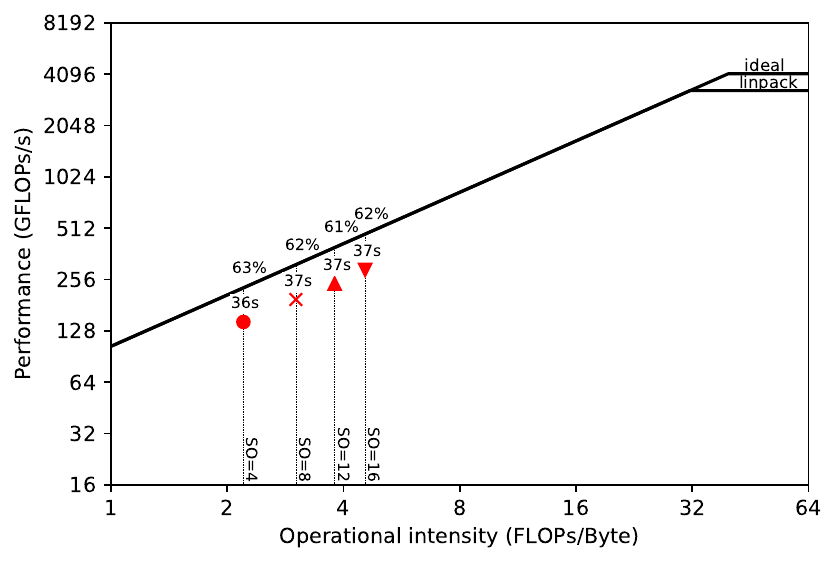}
\caption{\skl, \yask, $748^3$ grid points.}
\end{subfigure}%
\begin{subfigure}[b]{0.5\textwidth}
\includegraphics[width=\textwidth]{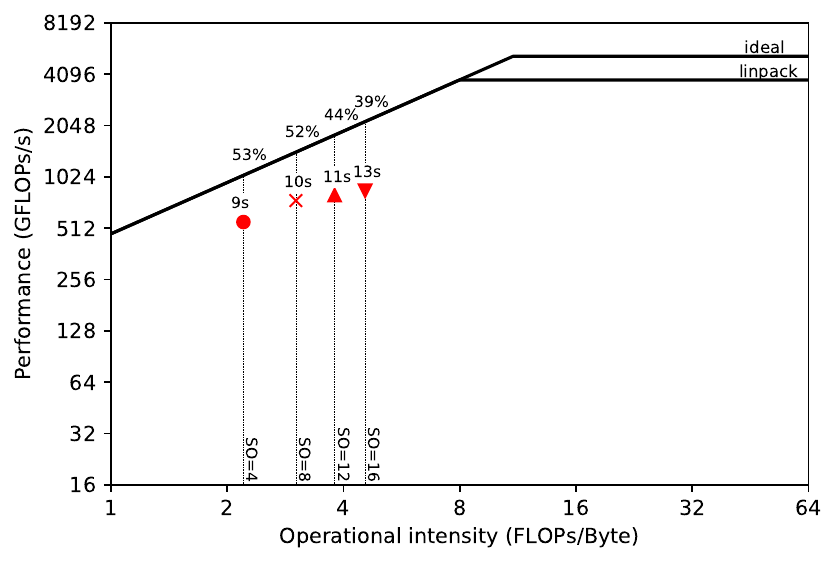}
\caption{\knl, \yask, $748^3$ grid points.}
\end{subfigure}%
\caption{Performance of \isotropic on multiple Devito backends and
    architectures.}
\label{fig:acoustic}
\end{figure}

The performance of \core on \skl, illustrated in
Figure~\ref{fig:acoustic:core-skl} (\yask uses slightly
smaller grids than \core due to a flaw in the API of \DevitoVer, which will
be fixed in \DevitoVerNext), degrades as the space order (henceforth, \so)
increases. In particular, it drops from 59$\%$ of the attainable machine peak
to 36$\%$ in the case of \so{=}16. This is the result of multiple issues. As
\so increases, the number of streams of unaligned virtual addresses also
increases, causing more pressure on the memory system. \vtune revealed that the
lack of split registers to efficiently handle split loads was a major source of
performance degradation. Another major issue for \isotropic on \core concerns
the quality of the generated SIMD code. The in-line vectorization performed by
the auto-vectorizer produces a large number of pack/unpack instructions to move
data between vector registers, which introduces substantial overhead. \vtune
also confirmed that, unsurprisingly, \isotropic is a memory-bound kernel.
Indeed, switching off the DSE basically did not impact the runtime, although it
did increase the operational intensity of the four test cases.

The performance of \core on \knl is not as good as that on \skl.
Figure~\ref{fig:acoustic:core-knl} shows an analogous trend to that on \skl,
with the attainable machine peak systematically dropping as \so increases. The
issue is that here the distance from the peak is even larger. This simply
suggests that \core is failing at exploiting the various levels of parallelism
available on \knl.

The \yask backend overcomes all major limitations to which \core is subjected.
On both \skl and \knl, \yask outperforms \core, essentially since it does not
suffer from the issues presented above. Vector folding reduces memory-read streams;
software prefetching helps especially for larger values of \so; and
hierarchical OpenMP parallelism is fundamental to leverage shared caches. The
speed-up on \knl is remarkable, since even in the best scenario for \core
(\so{=}4), \yask is roughly $3\times$ faster, and more than $4\times$ faster
when \so{=}12.

\subsection{Performance: acoustic wave in tilted transverse isotropy model}
\label{sec:performance:tti}
This sections illustrates the performance of \tti with the \core backend. \tti
cannot be run on the \yask backend in \DevitoVer as some fundamental features
are still missing; this is part of our future work (more details in
Section~\ref{sec:further-work}). 

\begin{figure}[htpb]
\centering
\begin{subfigure}[b]{0.5\textwidth}
\includegraphics[width=\textwidth]{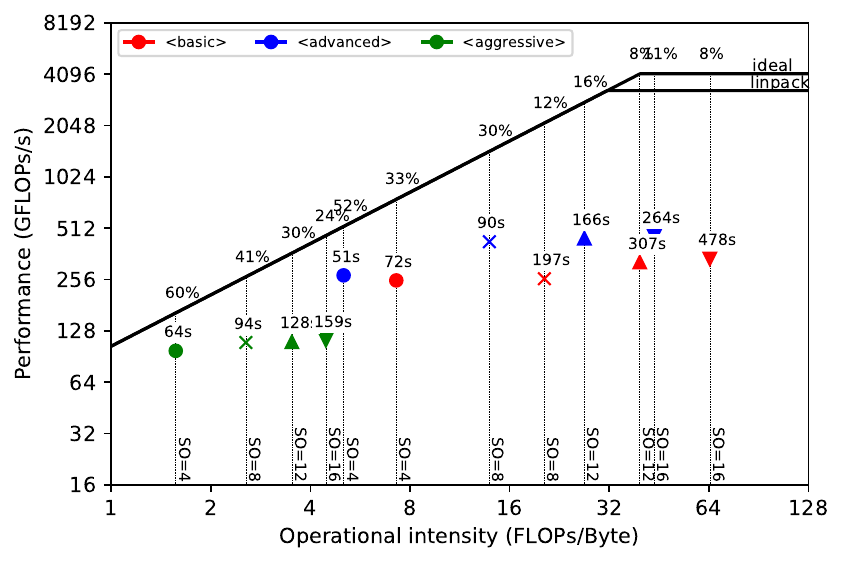}
\caption{\skl, $512^3$ grid points.}
\end{subfigure}%
\begin{subfigure}[b]{0.5\textwidth}
\includegraphics[width=\textwidth]{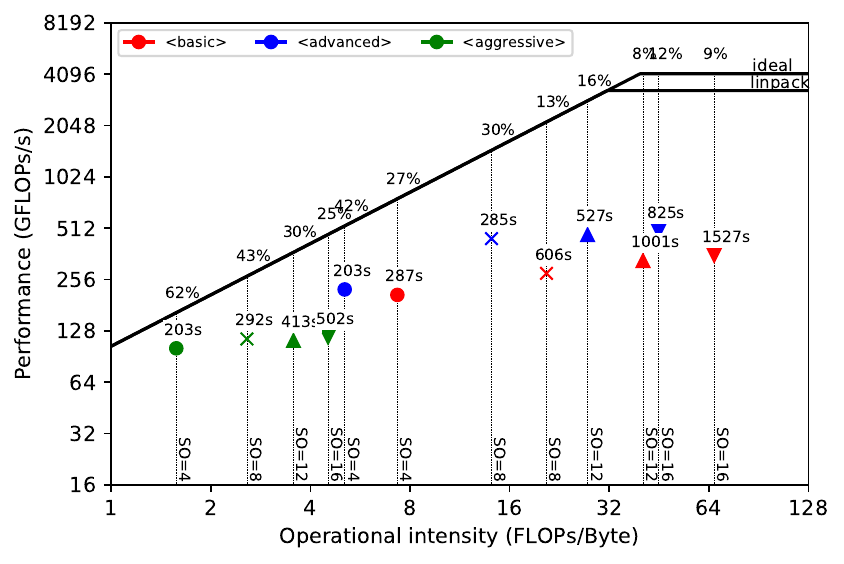}
\caption{\skl, $768^3$ grid points.}
\end{subfigure}%
~\\
\begin{subfigure}[b]{0.5\textwidth}
\includegraphics[width=\textwidth]{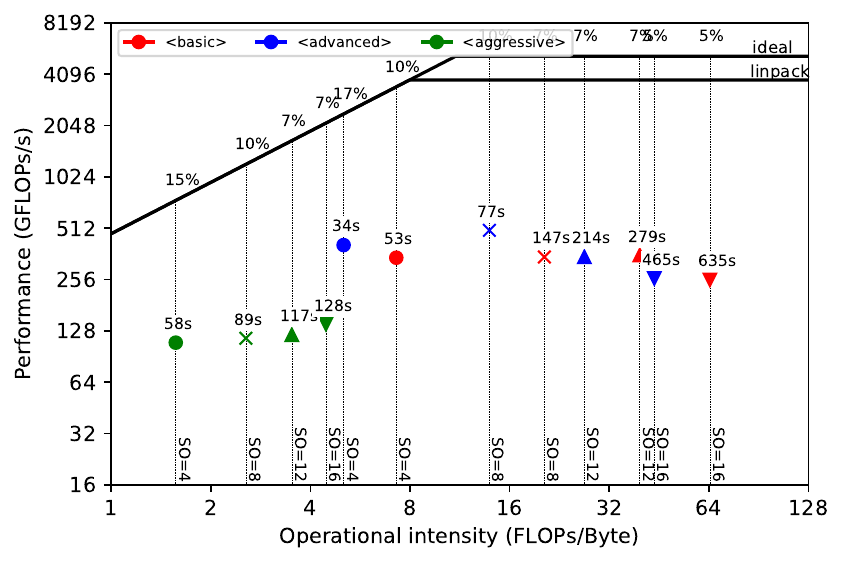}
\caption{\knl, $512^3$ grid points.}
\end{subfigure}%
\begin{subfigure}[b]{0.5\textwidth}
\includegraphics[width=\textwidth]{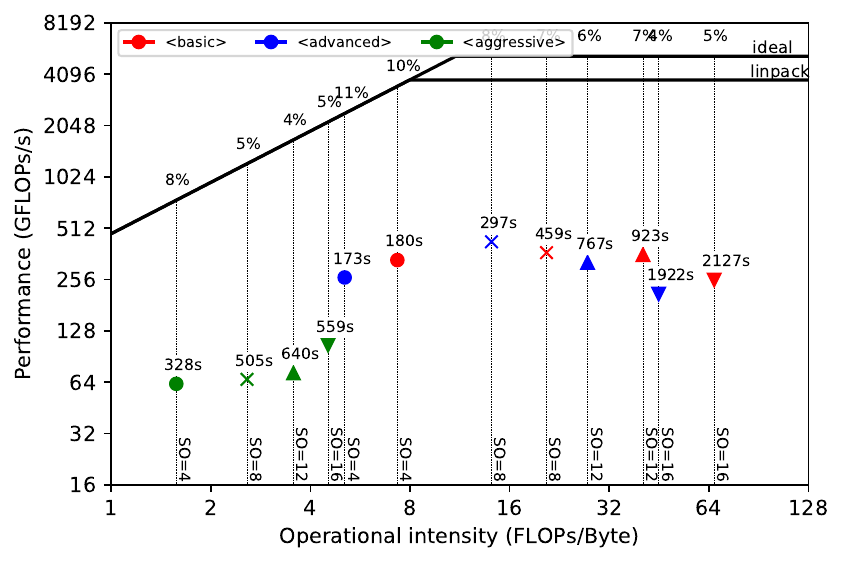}
\caption{\knl, $768^3$ grid points.}\label{fig:tti-knl-768}
\end{subfigure}%
\caption{Performance of \tti on \core for different architectures and grids.}
\label{fig:tti-perf}
\end{figure}

Unlike \isotropic, \tti significantly benefits from different levels of DSE
optimizations, which play a key role in reducing the operation count as well as
the register pressure. Figure~\ref{fig:tti-perf} displays the performance of
\tti for the usual range of space orders on \skl and \knl, for two different
cubic grids.

Generally, \tti does not reach the same level of performance as \isotropic.
This is not surprising given the complexity of the PDEs (e.g., in terms of
differential operators), which translates into code with much higher arithmetic
intensity. In \tti, the memory system is stressed by a considerably larger
number of loads per loop iteration than in \isotropic. On \skl, we ran
performance-profiling analyses using \vtune. We determined that the major issues are
pressure on both L1 cache (lack of split registers, insufficient ``fill
buffers'' to handle requests to the other levels of the hierarchy) and DRAM
(bandwidth and latency). Clearly, this is only a summary from some sample
kernels---the actual situation varies depending on the DSE optimizations as
well as the \so employed.

It is notable that on both \skl and \knl, and on both grids, the cutoff
point beyond which \optadv results in worse runtimes than \optagg is \so{=}8.
One issue with \optagg is that to avoid redundant computation, not only
additional memory is required, but also more data communication may occur
through caches, rather than through registers. In
Figure~\ref{lst:sie}, for example, we can easily deduce that {\tt temp}
is first stored, and then reloaded in the subsequent loop nest. This is an
overhead that \optadv does not pay, since temporaries are communicated through
registers, for as much as possible. Beyond \so{=}8, however, this overhead is
overtaken by the reduction in operation count, which grows almost quadratically
with \so, as reported in Table~\ref{table:tti:ops}.

\begin{table}[tbhp]
{\footnotesize
  \caption{Operation counts for different DSE modes in \tti}\label{table:tti:ops}
\begin{center}
  \begin{tabular}{|c|c|c|c|} \hline
      \so & \optbasic & \optadv & \optagg \\ \hline
      4 & 299 & 260 & 102 \\
      8 & 857 & 707 & 168 \\
      12 & 1703 & 1370 & 234 \\
      16 & 2837 & 2249 & 300 \\ \hline
  \end{tabular}
\end{center}
}
\end{table}

The performance on \knl is overall disappointing. This is unfortunately caused
by multiple factors---some of which already discussed in the previous
sections. These results, and more in general, the need for performance
portability across future (\Intel or non-\Intel) architectures, motivated the
ongoing \yask project. Here, the overarching issue is the inability to exploit
the multiple levels of parallelism typical of architectures such as \knl.
Approximately 17$\%$ of the attainable peak is obtained when \so{=}4 with
\optadv (best runtime out of the three DSE modes for the given space order).
This occurs when using $512^3$ points per grid, which allows the working set to
completely fit in MCDRAM (our calculations estimated a size of roughly 7.5GB).
With the larger grid size (Figure~\ref{fig:tti-knl-768}), the working set
increases up to 25.5GB, which exceeds the MCDRAM capacity. This partly
accounts for the 5$\times$ slow down in runtime (from 34s to 173s) in spite of
only a 3$\times$ increase in number of grid points computed per time iteration.

\subsection{Overhead summary}
To run an \Operator, there are four major sources of overhead:
\begin{description}
    \item[Code generation] The phase during which the high-level symbolic
        specification is lowered into C/C++.
    \item[Compilation into a shared object] The generated C/C++ file is
        compiled into a shared object by a C/C++ compiler with optimizations
        enabled. The time spent in this phase highly depends on the quality of
        the generated code.
    \item[Calling the shared object] This requires analyzing the user input,
        provided at \Operator application time, and forwarding it to the loaded
        shared object.
    \item[Auto-tuning] This step is optional. Its impact varies greatly across
        different problem sizes and even across backends (\core, \yask).
\end{description}

On \skl, Devito's turnaround times for all of these four phases are extremely
quick, even in the most complex problems in which hundreds of lines of code are
generated (e.g., high-order \tti). The \Intel compiler took less than 7 seconds
to build \tti\xspace \so{=}16 at the maximum optimization level, while the
\Operator required around 3 seconds to emit the C code (with DSE \optagg).
Calling the loaded shared object from \Python takes negligible time, despite
the extensive checks to validate the arguments. Auto-tuning took 3 minutes to
complete (heuristic-based search); however, from a user perspective, this is
hardly relevant as auto-tuning is disabled (by default) until production or
benchmark runs. All these times improves, even significantly, as the arithmetic
complexity of a problem decreases (e.g., at lower orders or when considering
\isotropic).

On \knl, due to weaker single-core performance, the overheads are more
pronounced. It took slightly more than one minute to produce a shared object
for \tti\xspace \so{=}16. Auto-tuning took around 15 minutes. Since it is unlikely
that a \knl will ever be used as a development platform, these overheads are
easily amortized out during production runs.

With the \yask backend the compilation times tend to increase, though the order
of magnitude is still the same as with \core. All other phases are substantially
unchanged.

For all experiments, we report the time spent in each of these phases in the
logs available at~\cite{devito-performance}.

\section{Further work}
\label{sec:further-work}

While many simulation and inversion problems such as full-waveform inversion
only require the solver to run on a single shared memory node, many other applications
require support for distributed memory parallelism (typically via MPI) so that
the solver can run across multiple compute nodes. The immediate plan is to leverage
\yask{'}s MPI support, and perhaps to include MPI support into \core at a later
stage.
Another important feature is staggered grids, which are necessary for a wide
range of FD discretization methods (e.g. modelling elastic wave propagation).
Basic support for staggered grids is already included in \DevitoVer, but
currently only through a low-level API---the principle
of graceful degradation in action. We plan to make the use of this feature
more convenient.

As discussed in Section~\ref{sec:performance:tti}, the \yask backend is not
feature-complete yet; in particular, it cannot run the \tti equations in the
presence of array temporaries. As \tti is among the most advanced models for
wave propagation used in industry, extending Devito in this direction has high
priority.

There also is a range of advanced performance optimization techniques that we
want to implement, such as ``time tiling'' (i.e., loop blocking across the time
dimension), on-the-fly data compression, and mixed-precision arithmetic
exploiting application knowledge.  Finally, there is an on-going effort towards
adding an \ops~\cite{OPS} backend, which will enable code generation for GPUs and
also supports distributed memory parallelism via MPI.

\section{Conclusions}
\label{sec:conclusions}
Devito is a system to automate high-performance stencil computations. While
Devito provides a \Python{-based} syntax to easily express FD approximations of
PDEs, it is not limited to finite differences. A Devito \Operator can implement
arbitrary loop nests, and can evaluate arbitrarily long sequences of
heterogeneous expressions such as those arising in FD solvers, linear algebra,
or interpolation. The compiler technology builds upon years of experience from
other DSL-based systems such as FEniCS and Firedrake, and wherever possible
Devito uses existing software components including \Sympy and \numpy, and YASK.
The experiments in this article show that Devito can generate production-level
code with compelling performance on state-of-the-art architectures.




%% file: symbolics.tex
The Devito DSL allows concise expression of \FD and general stencil operations
using a mathematical notation. It uses \Sympy~\cite{sympy-bib} for the
specification and manipulation of stencil expressions. In this section, we
describe the use of Devito's DSL to build PDE solvers.  Although the examples
used here are for \FD, the DSL can describe a large class of operations, such
as convolutions or basic linear algebra operations (e.g., chained tensor
multiplications).
\begin{figure}\centering
  \includegraphics[width=.80\textwidth]{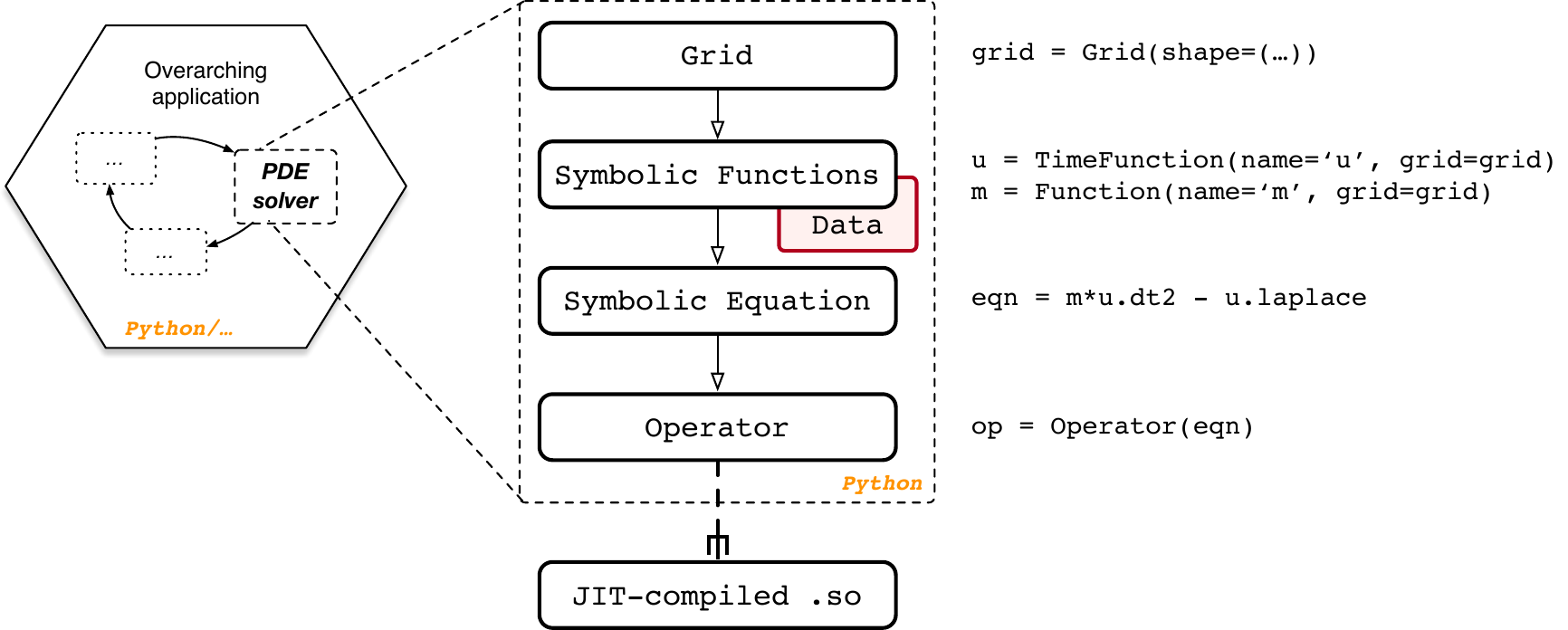}
  \caption{The typical usage of Devito within a larger application.}
  \label{fig:workflow}
\end{figure}

\subsection{Symbolic types}
\label{sec:api:symbolics}

The key steps to implement a numerical kernel with Devito are shown in
Figure~\ref{fig:workflow}. We describe this workflow, as well as fundamental
features of the Devito API, using the acoustic wave equation, also known as
d'Alembertian or Box operator. Its continuous form is given by:

\begin{align}\label{eq:WE}
\begin{aligned}
    {m}(x,y,z) \frac{d^2 {u}(x,y,z,t)}{dt^2} - \nabla^2 {u}(x,y,z,t) &= {q}_s, \\
    {u}(x,y,z,0) &= 0, \\
    \frac{d {u}(x,y,z,t)}{dt}|_{t=0} &= 0,
\end{aligned}
\end{align}
where the variables of this expression are defined as follows:
\begin{itemize}
  \item ${m}(x,y,z)=\frac{1}{c(x,y,z)^2}$, is the parametrization of the 
  	subsurface with $c(x,y,z)$ being the speed of sound as a function of the 
	three space coordinates $(x,y,z)$;

  \item ${u}(x,y,z,t)$, is the spatially varying acoustic wavefield, with the
       additional dimension of time $t$;

  \item ${q}_s$ is the source term, which is a point source in this case. 
\end{itemize}
The first step towards solving this equation is the definition of a discrete
computational grid on which the model parameters, wavefields, and source are
defined. The computational grid is defined as a \texttt{Grid(shape)} object,
where \shape is the number of grid points in each spatial dimension.  Optional
arguments for instantiating a \texttt{Grid} are \extent, which defines the
extent in physical units, and \origin, the origin of the coordinate system, with
respect to which all other coordinates are defined.

The next step is the symbolic definition of the squared slowness, wavefield,
and source. For this, we introduce some fundamental types.
\begin{itemize}
  \item \Function represents a discrete spatially varying function, such as
      the velocity. A \Function is instantiated for a defined \name and a given
        \Grid.

  \item \TimeFunction represents a discrete function that is both spatially
      varying and time dependent, such as wavefields. Again, a \TimeFunction
        object is defined on an existing \Grid and is identified by its \name.
  
  \item \SparseFunction and \SparseTimeFunction represent sparse functions,
      that is functions that are only defined over a subset of the grid, such
        as a seismic point source. The corresponding object is defined on a
        \Grid, identified by a \name, and also requires the \coordinates
        defining the location of the sparse points.
\end{itemize}

Apart from the grid information, these objects carry their respective \FD
discretization information in space and time. They also have a \data field that
contains values of the respective function at the defined grid points. By
default, \data is initialized with zeros and therefore automatically satisfies
the initial conditions from Equation~\ref{eq:WE}. The initialization of the
fields to solve the wave equation over a one-dimensional grid is displayed in
Listing~\ref{lst:wave-setup}.

\begin{listing}
\begin{lstlisting}
>>> from devito import Grid, TimeFunction, Function, SparseTimeFunction
>>> g = Grid(shape=(nx,), origin=(ox,), extent=(sx,))
>>> u = TimeFunction(name="u", grid=g, space_order=2, time_order=2)  # Wavefield
>>> m = Function(name="m",  grid=g)  # Physical parameter
>>> q = SparseTimeFunction(name="q", grid=g, coordinates=coordinates)  # Source
\end{lstlisting}
\caption{Setup \Function{s} to express and solve the acoustic wave equation.}
    \label{lst:wave-setup}
\end{listing}

\subsection{Discretization}

With symbolic objects that represent the discrete velocity model, wavefields,
and source function, we can now define the full discretized wave equation. As
mentioned earlier, one of the main features of Devito is the possibility to
formulate stencil computations as concise mathematical expressions. To do so,
we provide shortcuts to classic \FD stencils, as well as the functions to
define arbitrary stencils. The shortcuts are accessed as object properties and
are supported by \TimeFunction and \Function objects. For example, we can take
spatial and temporal derivatives of the wavefield \texttt{u} via the shorthand
expressions \texttt{u.dx} and \texttt{u.dt} (Listing~\ref{lst:fd-1}).

\begin{listing}
\begin{lstlisting}
>>> u.dx
-u(t, x - h_x)/(2*h_x) + u(t, x + h_x)/(2*h_x)
>>> u.dt
-u(t - dt, x)/(2*dt) + u(t + dt, x)/(2*dt)
>>> u.dt2
-2*u(t, x)/dt**2 + u(t - dt, x)/dt**2 + u(t + dt, x)/dt**2
\end{lstlisting}
\caption{Example of spatial and temporal \FD stencil creation.}
\label{lst:fd-1}
\end{listing}

Furthermore, Devito provides shortcuts for common differential operations such
as the Laplacian via \texttt{u.laplace}. The full discrete wave equation can
then be implemented in a single line of \Python (Listing~\ref{lst:fd-2}).

\begin{listing}
\begin{lstlisting}
>>> wave_equation = m * u.dt2 - u.laplace
>>> wave_equation
(-2*u(t, x)/dt**2 + u(t - dt, x)/dt**2 + u(t + dt, x)/dt**2)*m(x) + 2*u(t, x)/h_x**2 - u(t, x - h_x)/h_x**2 - u(t, x + h_x)/h_x**2
\end{lstlisting}
\caption{Expressing the wave equation.}
\label{lst:fd-2}
\end{listing}

To solve the time-dependent wave equation with an explicit time-stepping
scheme, the symbolic expression representing our PDE has to be rearranged
such that it yields an update rule for the wavefield $u$ at the next time
step: $u(t+dt) = f(u(t), u(t-dt))$). Devito allows to rearrange the PDE
expression automatically using the \texttt{solve} function, as shown in
Listing~\ref{lst:fd-3}.

\begin{listing}[H]
\begin{lstlisting}
>>> from devito import Eq, INTERIOR, solve
>>> stencil = Eq(u.forward, solve(wave_equation, u.forward), region=INTERIOR)
>>> stencil
Eq(u(t + dt, x), -2*dt**2*u(t, x)/(h_x**2*m(x)) + dt**2*u(t, x - h_x)/(h_x**2*m(x)) + dt**2*u(t, x + h_x)/(h_x**2*m(x)) + 2*u(t, x) - u(t - dt, x))
\end{lstlisting}
\caption{Time-stepping scheme for the acoustic wave equation. {\tt
region=INTERIOR} ensures that the Dirichlet boundary conditions at the
edges of the Grid are satisfied.}
\label{lst:fd-3}
\end{listing}

Note that the \texttt{stencil} expression in Listing~\ref{lst:fd-3} does not
yet contain the point source $q$. This could be included as a regular \Function
which has zeros all over the grid except for a few points, but it would
obviously be wasteful. Instead, \SparseFunction{s} allow to perform operations,
such as injecting a source or sampling the wavefield, at a subset of points
determined by \coordinates. In general, receivers (where the solution
is to sampled) are not co-located with grid points. Therefore, an interpolation
operator is needed (e.g.  trilinear interpolation for 3D). To ensure a
consistent discrete adjoint, source terms are implemented as the adjoint of the
interpolation operator used, that is the gather operation for interpolation
becomes a scatter operation for source injection.  Equation~\ref{eq:interp}
gives the expressions for linear interpolation in 1D assuming the origin is zero
for readability.

\begin{align}\label{eq:interp}
\begin{aligned}
    \text{Find the two closest indices:\ \ \ }& x_1 = \floor[\Big]{\frac{q_{coords}[i]}{h_x}}, \ x_2 = x_1 + 1 \\
    \text{Interpolation coefficients:\ \ \ }& c_1 = 1 - \frac{q_{coords}[i] - x_1}{x_2 - x_1}, \ c_2 =  1 - \frac{x_2 - q_{coords}[i]}{x_2 - x_1} \\
    \text{Interpolate:\ \ \ }& q[i] = c_1 * u[t, x_1] + c_2 * u[t, x_2] \\
    \text{Inject:\ \ \ }& u[t, x_1] = q[i] * c_1, \ u[t, x_2] = q[i] * c_2
\end{aligned}
\end{align}

To inject a point source defined at the physical location \texttt{q\_coords} into the
\texttt{stencil} expression, we use the \texttt{inject} function of the
\SparseTimeFunction object that represents our seismic source
(Listing~\ref{lst:finject}).\footnote{More complicated interpolation schemes
can be defined by precomputing the grid points corresponding to each sparse
point, and their respective coefficients. The result can then be used to create
a \PrecomputedSparseFunction, which behaves like a \SparseFunction at the
symbolic level.}

\begin{listing}
\begin{lstlisting}
>>> injection = q.inject(field=u.forward, expr=dt**2 * q / m)
>>> injection 
[Eq(u[t + 1, INT(floor((-o_x + q_coords[p_q, 0])/h_x))], dt**2*(1 - FLOAT(-h_x*INT(floor((-o_x + q_coords[p_q, 0])/h_x)) - o_x + q_coords[p_q, 0])/h_x)*q[time, p_q]/m[INT(floor((-o_x + q_coords[p_q, 0])/h_x))] + u[t + 1, INT(floor((-o_x + q_coords[p_q, 0])/h_x))]),
 Eq(u[t + 1, INT(floor((-o_x + q_coords[p_q, 0])/h_x)) + 1], dt**2*FLOAT(-h_x*INT(floor((-o_x + q_coords[p_q, 0])/h_x)) - o_x + q_coords[p_q, 0])*q[time, p_q]/(h_x*m[INT(floor((-o_x + q_coords[p_q, 0])/h_x)) + 1]) + u[t + 1, INT(floor((-o_x + q_coords[p_q, 0])/h_x)) + 1])]
\end{lstlisting}
\caption{Expressing the injection of a source into a field.}
\label{lst:finject}
\end{listing}

The \texttt{inject} function takes the field being updated as an input argument
(in this case \texttt{u.forward}), while \texttt{expr=dt**2 * q / m} is the
expression being injected. The result of the \texttt{inject} function is a list
of symbolic expressions that correspond to the different steps of
Equation~\ref{eq:interp}. As we shall see, these expressions are eventually
joined together and used to create an \Operator object---the solver of our
PDE.

\subsection{Boundary conditions}
Simple boundary conditions (BCs), such as Dirichlet BCs, can be imposed on
individual equations through special keywords (see Listing~\ref{lst:fd-3}).
For more exotic schemes, instead, the BCs need to be explicitly written (e.g.,
Higdon BCs~\cite{Higdon-bcs}), just like any of the symbolic expressions
defined in the Listings above. For reasons of space, this aspect is not
elaborated further; the interested reader may refer to~\cite{notebook-bcs}.


\subsection{Control flow}
By default, the extent of a \TimeFunction in the time dimension is limited by
its time order. Hence, the shape of $u$ in Listing~\ref{lst:wave-setup} is
$(time\_order + 1, nx) = (3, nx)$. The iterative method will then access $u$ via
modulo iteration, that is $u[t \% 3, ...]$. In many scenarios, however, the
entire time history, or at least periodic time slices, should be saved (e.g.,
for inversion algorithms). Listing~\ref{lst:fd-4} expands our running example
with an equation that saves the content of $u$ every $4$ iterations, up to a
maximum of $save=100$ time slices.

\begin{listing}[H]
\begin{lstlisting}
>>> from devito import ConditionalDimension
>>> ts = ConditionalDimension('ts', parent=g.time_dim, factor=4)
>>> us = TimeFunction(name='us', grid=g, save=100, time_dim=ts)
>>> save = Eq(us, u)
\end{lstlisting}
\caption{Implementation of time sub-sampling.}
\label{lst:fd-4}
\end{listing}

In general, all equations that access \Function{s} (or \TimeFunction{s})
employing one or more \ConditionalDimension{s} will be conditionally executed.
The condition may be a number indicating how many iterations should pass
between two executions of the same equation, or even an arbitrarily complex
expression.

\subsection{Domain, halo, and padding regions}
\label{sec:api:domain-halo}

A \Function internally distinguishes between three regions of points.
\begin{description}
\item[Domain] Represents the {\it computational domain} of the \Function and is 
inferred from the input \Grid. This includes any elements added to the
{\it physical domain} purely for computational purposes,
e.g. absorbing boundary layers. 
\item[Halo] The grid points surrounding the domain region, i.e., ``ghost''
points that are accessed by the stencil when iterating in proximity
of the domain boundary. 
\item[Padding] The grid points surrounding the halo region, which are allocated
for performance optimizations, such as data alignment. Normally this
region should be of no interest to a user of Devito, except for
precise measurement of memory allocated for each \Function.
\end{description}

%% file: yask.tex
\begin{figure}[htb]
\small
\begin{center}
\begin{tabulary}{\textwidth}{ C C C }
\includegraphics[width=4.4cm,keepaspectratio]{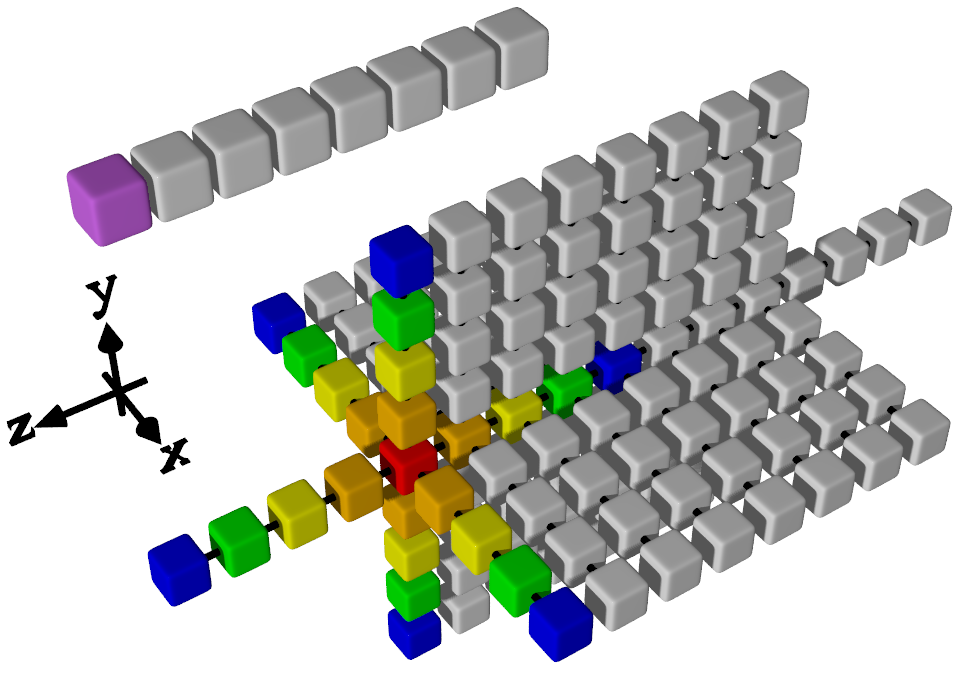} &
\includegraphics[width=3.4cm,keepaspectratio]{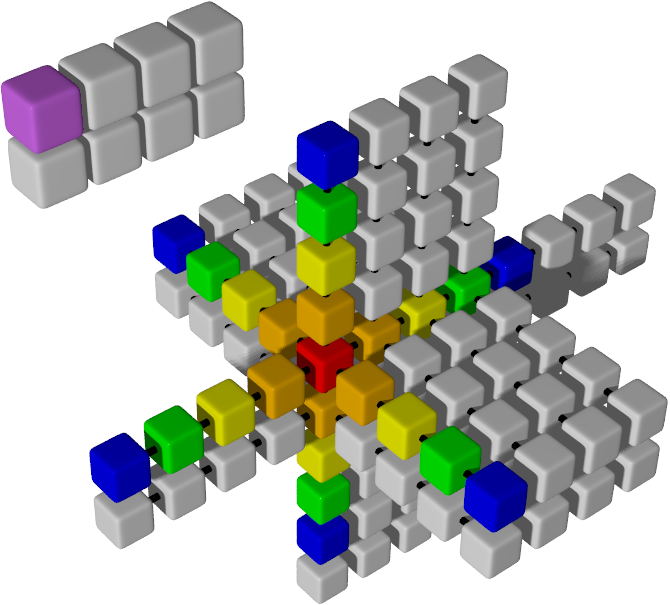} &
\includegraphics[width=3.4cm,keepaspectratio]{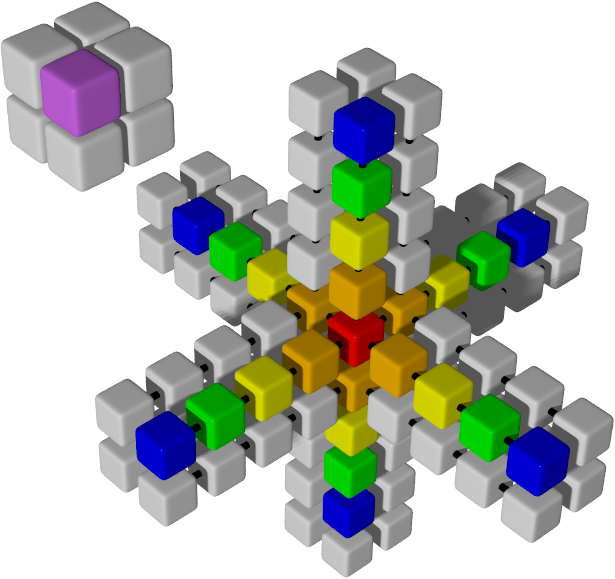} \\
a. $1\times 1 \times 8$ 1D fold &
b. $1\times 2 \times 4$ 2D fold & 
c. $2\times 2 \times 2$ 3D fold
\end{tabulary}
\end{center}
\caption{Various folds of 8 elements~\cite{yask-vec-folding}. 
The smaller diagram in the upper-left of each sub-figure illustrates a single SIMD layout, which is also the configuration of the output elements from a single SIMD computation.
The larger diagram shows the SIMD input values needed for a typical 25-point stencil, e.g., from an 8th-order finite-difference approximation of an isotropic acoustic wave. 
The colored elements highlight the first element in the output layout (purple element) and the corresponding elements in the inputs (red through blue elements, where the different colors indicate the distance from the center).
Note that the $1\times 1 \times 8$ 1D fold corresponds to the traditional in-line vectorization.}
\label{fig:folds}
\end{figure}

YASK---Yet Another Stencil Kit\footnote{Formerly, Yet Another Stencil Kernel.}---is an open-source C++ software framework for generating high-performance implementations of stencil codes for \Intel\xspace \Xeon and \XeonPhi processors.
Previous publications on YASK have discussed its overall structure~\cite{yask-main} and its application to the \XeonPhi x100 family (code-named Knights Corner)~\cite{yask-vec-folding} and \XeonPhi x200 family (code-named Knights Landing)~\cite{hbm-tiling-2016,seismic-sims-on-knl:tobin-isc17,elastic-wave-xeon-phi-2018} many-core CPUs.
Unlike Devito, it does not expose a symbolic language to the programmer or create stencils from finite-difference approximations of differential equations.
Rather, the programmer provides simple declarative descriptions of the stencil equations using a C++ or Python API.
Thus, Devito operates at a level of abstraction higher than that of YASK, while YASK provides performance portability across Intel architectures and is more focused on low-level optimizations. 
Following is a sample of some of the optimizations provided by YASK:
\begin{itemize}
    \item {\bf Vector-folding.}
In traditional SIMD vectorization, such as that provided by an auto-vectorizing compiler, the vector elements are arranged sequentially along the unit-stride dimension of the grid, which is also the dimension iterated over in the inner-most loop of the stencil kernel.
Vector-folding is an alternative data-layout method whereby neighboring elements are arranged in small \emph{multi-dimensional} tiles.
Figure~\ref{fig:folds} illustrates three ways to pack eight double-precision floating-point values into a 512-bit SIMD register.  
Figure~\ref{fig:folds}a shows a traditional 1D ``in-line'' layout, and \ref{fig:folds}b and \ref{fig:folds}c show alternative 2D and 3D ``folded'' layouts.
Furthermore, these tiles may be ordered in memory in a dimension independent of the dimensions used in vectorization~\cite{yask-vec-folding}.
The combination of these two techniques can significantly increase overlap and reuse between successive stencil-application iterations, reducing the memory-bandwidth demand.
For stencils that are bandwidth-bound, this can provide significant performance gains~\cite{yask-vec-folding,seismic-sims-on-knl:tobin-isc17}.
\item {\bf Software prefetching.} Many high-order or staggered-grid stencils require multiple streams of data to be read from memory, which can overwhelm the hardware prefetchers.
YASK can automatically generate software-prefetch instructions to improve the cache hit rates, especially on Xeon Phi CPUs.
\item {\bf Hierarchical parallelism.} Dividing the spatial domain into tiles to increase temporal cache locality is a common stencil optimization as discussed earlier. 
When implementing this technique, sometimes called ``cache-blocking'', it is typical to assign each thread to one or more small rectilinear subsets of the domain in which to apply the stencil(s).
However, if these threads share caches, one thread's data will often evict data needed later by another thread, reducing the effective capacity of the cache.
YASK addresses this by employing two levels of OpenMP parallelization: the outer level of parallel loops are applied across the cache-blocks, and an inner level is applied across sub-blocks within those tiles.
In the case of the Xeon Phi, the eight hyper-threads that share each L2 cache can now cooperate on filling and reusing the data in the cache, rather than evicting each other's data.


\end{itemize}

YASK also provides other optimizations, such as temporal tiling and MPI support that are not exploited by Devito at the time of writing.
The interested reader may refer to~\cite{hbm-tiling-2016,stencil-tiling-fgcs-2019}.

To leverage both the symbolic processing of Devito and the low-level optimizations of YASK, 
we have integrated the YASK framework into
the Devito package. In essence, the Devito \yask backend exploits the
intermediate representation of an \Operator to generate YASK kernels. 
In \DevitoVer, roughly
70$\%$ of the Devito API is supported by the \yask backend.\footnote{At the time
of writing, reaching feature-completeness is one the major on-going development
efforts.}

